\documentclass[final,authoryear,3p,times,twocolumn]{elsarticle}
\usepackage[latin2]{inputenc}
\usepackage{graphicx}
\usepackage{amssymb}
\usepackage{amsthm}
\usepackage{ulem}

\journal{Applied Radiation and Isotopes 113(2016)96}

\begin{document}

\begin{frontmatter}



\title{Activation cross sections of proton induced nuclear reactions on gold up to 65 MeV}


\author[1]{F. Ditr\'oi\corref{*}}
\author[1]{F. T\'ark\'anyi}
\author[1]{S. Tak\'acs}
%\author[1] {J. Csikai}
\author[2]{A. Hermanne}
%\author[3]{S. Uddin}
%\author[3]{M. Baba}
%\author[4]{A.V. Ignatyuk}

\cortext[*]{Corresponding author: ditroi@atomki.hu}

\address[1]{Institute for Nuclear Research, Hungarian Academy of Sciences (ATOMKI),  Debrecen, Hungary}
\address[2]{Cyclotron Laboratory, Vrije Universiteit Brussel (VUB), Brussels, Belgium}
\address[4]{Institute of Physics and Power Engineering (IPPE), Obninsk, Russia}
\address[3]{Cyclotron Radioisotope Center (CYRIC), Tohoku University, Sendai, Japan}

\begin{abstract}
\noindent Activation cross sections of proton induced reactions on gold for production of $^{197m,197g,195m,195g, 193m,193g,192}$Hg, $^{196m,196g(cum),195g(cum),194,191(cum)}$Au, $^{191(cum)}$Pt and $^{192}$Ir were measured up to 65 MeV proton energy, some of them for the first time. The new data are in acceptably good agreement with the recently published earlier experimental data in the overlapping energy region. The experimental data are compared with the predictions of the TALYS 1.6 (results in TENDL-2015 on-line library) and EMPIRE 3.2 code.
\end{abstract}

\begin{keyword}
natural gold target\sep activation cross section\sep Hg, Au, Pt and Ir radioisotopes\sep physical yield\sep medical and industrial applications

\end{keyword}

\end{frontmatter}

% \linenumbers

\section{Introduction}
\label{1}
The experimental proton and deuteron activation cross section data on gold are important in different application fields. Out of the reaction products the radionuclides of mercury are used in monitoring the distribution and the accumulation of mercury in different parts of the body and to study mercury transformations in environmental systems. The accelerator-produced isomers are broadly used in diagnostic medical applications: $^{197m,g}$Hg appear to be new potential candidates for therapy. For investigation of specimens containing metallic gold and gold alloys, the longer-lived $^{198g}$Au (T$_{1/2}$ = 2.7 d) and $^{196g}$Au (T$_{1/2}$ = 6.2 d) are suitable. Furthermore, experimental studies of the isomers and comparison of the results with nuclear reaction model calculations significantly assist to further development of nuclear reaction theories. We earlier published results of investigations on gold activation cross sections where the different applications were discussed in more detail \citep{IAEA-NDS, Szelecsenyi1996, Szelecsenyi1997, Tarkanyi2011, Tarkanyi2015}.
The present data are produced in the frame of a systematic study of excitation functions of light charged particles. Our earlier investigation on proton induced activation on gold was limited to 30 MeV incident particle energy \citep{Szelecsenyi1996}. This recent investigation is dedicated mainly to the determination of excitation functions on gallium up to 65 MeV and $^{64}$Ni up to 30 MeV \citep{Amjed, Hermanne}, we have used Au-foils as backing of electrodeposited GaNi and $^{64}$Ni targets. As during the evaluation of the gamma spectra we got valuable information on activation cross sections also on gold, we decided to share the new information with the public.

\section{Earlier experimental data}
\label{2}

The main goal of the present work was to extend the energy range of the experimental data for cross sections of radioisotopes produced from gold by proton activation. Besides this goal we could also clarify the discrepancies occur between the relatively large number of data sets from different laboratories, as well as provide further input data for nuclear reaction model code development.  Due to different factors (monoisotopic target, easy target preparation, importance of proton activation data for different applications, easy measurable isomeric ratios, etc.) there is a relatively large set of experimental data available in the literature on proton induced activation cross sections on gold. When one overviews the years of the previous measurements, target preparation methods, irradiation circumstances, beam current measurements methods (monitoring), $\gamma$-spectrometry tools and the overall result parameters (in the last column of Table 1), one can have a preliminary impression about the quality of the particular data sets. This information can later be used by judging the goodness of the agreements. We collected the results related to the present study in Table 1.

\begin{table*}[t]
\tiny
\caption{Earlier experimental data}
\begin{center}

\begin{tabular}{|p{0.8in}|p{0.7in}|p{0.7in}|p{1.0in}|p{0.8in}|p{1.9in}|} \hline 
\textbf{Author} & \textbf{Target} & \textbf{Irradiation} & \textbf{Beam current\newline measurement\newline and monitor\newline reaction} & \textbf{Measurement \newline of activity} & \textbf{Reaction\newline Measured quantity\newline Number of measured data points \newline Energy range (MeV) } \\ \hline 
\citep{Boehm} & Au foil & Cyclotron\newline Single target &  & GM counter & ${}^{197m}$Hg/${}^{197g}$Hg, $\sigma$ rat, No.1,6.7 MeV \\ \hline 
\citep{Sosniak} & Au foil\newline 10 and 15 $\mu$m & Syncro-cyclotron & $^{12}$C(p,pn)$^{11}$C\newline ${}^{19}$F(p,pn)${}^{18}$F & Chemical separation \newline Kx, Scint. counter\newline  & ${}^{197}$Au(p,pn)${}^{196}$Au,$\sigma$, No. 5, 82-426~MeV~ \\ \hline 
\citep{Poffe} & Au foil & Syncro-cyclotron\newline Single target & ${}^{27}$Al(p,x)${}^{24}$Na\newline  & Chemical separation \newline Na(I) scintillator & ${}^{197}$Au(p,3n)${}^{195g}$Hg(m+),$\sigma$, No. 1, 155 MeV\newline $^{197}$Au(p,3n)${}^{197g}$Hg(m+),$\sigma$, No. 1, 155 MeV \\ \hline 
\citep{Yule} & Au foils 10-25 mg/cm${}^{2}$ & Syncro-cyclotron\newline Single target\newline (recoil catchers) & ${}^{27}$Al(p,x)${}^{24}$Na\newline ${}^{12}$C(p,pn)${}^{11}$C & Chemical separation \newline Prop. Counter \newline Scint. counter & $^{197}$Au(p,pn)${}^{196}$Au,s, No. 6, 17.7-31.8 MeV \\ \hline 
\citep{Vandenbosch} & Au foils\newline 100-600 $\mu$m & Cyclotron\newline Stacked target\newline  & Faraday cup & No Chemical separation\newline Na(I) scintillator & ${}^{197}$Au(p,n)$^{197m}$Hg,$^{197g}$Hg, $\sigma$ and $\sigma$${}_{rat}$, No.5,7.3-10.4 MeV \\ \hline 
\citep{Gusakow1960} & Au foil & Syncro-cyclotron\newline Single target & $^{27}$Al(p,x)${}^{24}$Na\newline $^{12}$C(p,pn)${}^{11}$C &  & ${}^{197}$Au(p,p2n)${}^{195g}$Au,$\sigma$, No. 9, 35.9-154.8 MeV \\ \hline 
\citep{Gusakow1961} & Au foil & Syncro-cyclotron\newline Single target & ${}^{27}$Al(p,x)$^{24}$Na\newline $^{12}$C(p,pn)${}^{11}$C & Chemical separation \newline $\gamma$-Na(I) scintillator & ${}^{197}$Au(p,pn)${}^{196}$Au,$\sigma$, No. 8, 40.2-155.5 MeV\newline $^{197}$Au(p,3n)${}^{195}$Hg,$\sigma$, No. 2, 60-80 MeV\newline ${}^{197}$Au(p,5n)${}^{193}$Hg,$\sigma$,No. 18, 40.4-155 MeV\newline ${}^{197}$Au(p,6n)${}^{192}$Hg,$\sigma$, No. 2, 60-80 MeV\newline ${}^{197}$Au(p,7n)${}^{191}$Hg,$\sigma$, No. 2, 60-80 MeV\newline ${}^{197}$Au(p,7n)$^{190}$Hg,$\sigma$, No. 1, 80 MeV \\ \hline 
\citep{Poffe} & Au foil & Syncro-cyclotron\newline Single target & ${}^{27}$Al(p.x)${}^{24}$Na\newline  & Chemical separation \newline Na(I) scintillator & ${}^{197}$Au(p,3n)${}^{195g}$Hg(m+),$\sigma$, No. 1, 155 MeV \\ \hline 
\citep{Kavanagh} & Au foil \newline 20-40 mg/cm${}^{2}$ & Syncro-cyclotron\newline Single target & ${}^{27}$Al(p.x)${}^{24}$Na\newline  & Chemical separation\newline Na(I) scintillator & ${}^{197}$Au(p,pn)${}^{196}$Au,$\sigma$, No. 26, 18-86 MeV\newline ${}^{197}$Au(p,p2n)${}^{195}$Au,$\sigma$, No. 19, 23.2-86 MeV\newline ${}^{197}$Au(p,p3n)${}^{194}$Au,$\sigma$, No. 20, 32.4-86 MeV \\ \hline 
(Hansen et al., 1962) & Au foil\newline 10 mg/cm${}^{2}$ & Cyclotron\newline  stacked foil & Faraday cup & Na(I) scintillator & ${}^{197}$Au(p,n)${}^{197}$Hg,$\sigma$, No. 8, 7-12.7$\sigma$ MeV\newline ${}^{197}$Au(p,n)${}^{197g}$Hg,$\sigma$, No. 8, 7-12.7 MeV\newline ${}^{197}$Au(p,n)${}^{197m}$Hg,$\sigma$, No. 8, 7-12.7 MeV \\ \hline 
\citep{Albouy} & Au foil & Syncro-cyclotron\newline Single target & ${}^{27}$Al(p.x)${}^{24}$Na\newline ${}^{12}$C(p,pn)${}^{11}$C\newline ${}^{12}$C(p,pn)${}^{7}$Be & Chemical separation\newline Prop. Count & ${}^{197}$Au(p,x)${}^{192}$Ir,$\sigma$, No. 9, 44.53-149.66 MeV \\ \hline 
\citep{Gritsyna} & Au foil\newline 5-30 mg/cm${}^{2}$ & LINAC\newline Stacked target &  & Na(I) & ${}^{197}$Au(p,n)${}^{197g}$Hg,$\sigma$, No. 13, 8.3-19.5 MeV\newline ${}^{197}$Au(p,n)${}^{197m}$Hg,$\sigma$, No. 13 , 8.3-19.5 MeV \\ \hline 
\citep{Thomas} & Au foil\newline 2 mg/cm${}^{2}$ & Cyclotron \newline Single target & Faraday cup\newline PP ion chamber\newline Energy range of scattered particles & Neutron\newline Liquid scintillator & ${}^{197}$Au(p,2n)${}^{196}$Hg,$\sigma$,No. 10, 9.7-13.9 MeV\newline ${}^{197}$Au(p,n)${}^{196}$Hg+${}^{197}$Au(p,pn)${}^{196}$Au,$\sigma$,No. 14, 7.6-13.9 MeV\newline  \\ \hline 
\citep{Chodil} & Au foil\newline 2 mg/cm${}^{2}$ mg/cm2). Au (2.0 mg/cm2). & Cyclotron \newline Single target & Faraday cup\newline PP ion chamber\newline Energy range of scattered particles & Neutron\newline Liquid scintillator & ${}^{197}$Au(p,2n)${}^{196}$Hg,$\sigma$, No. 15, 7-15 MeV\newline ${}^{197}$Au(p,n)${}^{196}$Hg+${}^{197}$Au(p,pn)${}^{196}$Au,$\sigma$, No. 15, 7-15 MeV\newline ${}^{197}$Au(p,x)n, $\sigma$, No. 20, 7-15 MeV\newline  \\ \hline 
\citep{Kusch} & Au foil & Linear accelerator\newline Stacked foil &  & Ge(Li) & ${}^{197}$Au(p,n)${}^{197m}$Hg,${}^{197g}$Hg, $\sigma$ and $\sigma$${}_{rat}$, No.6,6.5-9.5 MeV\newline ${}^{197}$Au(p,n)${}^{197}$Hg,$\sigma$, No. 4, 6.5-9.5 MeV \\ \hline 
\citep{Alderliesten} & Au foil & Cyclotron \newline  &  & Ge(Li) &  \\ \hline 
\citep{Probst} &  & Cyclotron \newline Single target & Faraday cup\newline  & Ge(Li) & ${}^{197}$Au(p,n)${}^{197g}$Hg,$\sigma$, No. 10, 7.3-17.9 MeV\newline ${}^{197}$Au(p,n)${}^{197m}$Hg,$\sigma$, No. 31 , 6.4-44.6 MeV\newline ${}^{197}$Au(p,3n)${}^{195g}$Hg,$\sigma$, No. 20, 17.9-44.6 MeV\newline ${}^{197}$Au(p,3n)${}^{195m}$Hg,$\sigma$, No. 20, 17.9-44.6 MeV\newline ${}^{197}$Au(p,3n)${}^{194}$Hg,$\sigma$, No. 10, 26.9-44.6 MeV\newline ${}^{197}$Au(p,5n)${}^{193g}$Hg,$\sigma$, No. 5, 36.1-44.9 MeV\newline ${}^{197}$Au(p,5n)${}^{193m}$Hg,$\sigma$, No. 5, 36.1-44.9 MeV \\ \hline 
\citep{Birattari} & Au foil\newline 26.23 mg/cm$^{2}$ & cyclotron & Faraday cup & Ge(Li) & $^{197}$Au(p,n)${}^{197g}$Hg,tty, No. 9, 7.4-20.7 MeV\newline ${}^{197}$Au(p,n)${}^{197m}$Hg,tty, No. 12 , 8.11-33.9 MeV\newline $^{197}$Au(p,3n)$^{195g}$Hg,tty, No. 10, 18.7-37 MeV\newline ${}^{197}$Au(p,3n)${}^{195m}$Hg,tty, No. 11, 18.8-43.4 MeV\newline ${}^{197}$Au(p,5n)$^{193m}$Hg,tty, No. 10, 36.4-44.6 MeV\newline ${}^{197}$Au(p,2n)${}^{196g}$Hg,tty, No. 19, 15.2-42.2 MeV\newline ${}^{197}$Au(p,2n)${}^{196m2}$Hg,tty, No. 9, 24-42.2 MeV\newline ${}^{197}$Au(p,2n)${}^{194}$Hg,tty, No. 5, 30.2-42.2 MeV \\ \hline 
\citep{Dmitriev1981} &  & Cyclotron \newline Stacked target &  & Ge(Li) & ${}^{197}$Au(p,pn)${}^{196}$Au,tty, No. 1, 22 MeV\newline ${}^{197}$Au(p,n)${}^{197g}$Hg,tty, No. 1, 22 MeV \\ \hline 
(Dmitriev, 1983) &  & Cyclotron \newline Stacked target &  & Ge(Li) & ${}^{197}$Au(p,n)${}^{197g}$Hg,py, No. 1, 22 MeV\newline ${}^{197}$Au(p,x)${}^{196}$Au,py, No. 1, 22 MeV \\ \hline 
\citep{Abe} &  & Cyclotron & $^{65}$Cu(p,n)$^{65}$Zn\newline  & Ge(Li)\newline  & ${}^{197}$Au(p,pn)${}^{196g}$Au(m+),tty, No. 1, 16 MeV\newline ${}^{197}$Au(p,pn)${}^{197g}$Hg(m+),tty, No. 1, 16 MeV\newline ${}^{197}$Au(p,pn)${}^{197g}$Hg,tty, No. 1, 16 MeV\newline ${}^{197}$Au(p,pn)${}^{197m}$Hg,tty, No. 1, 16 MeV \\ \hline 
\citep{Bonardi1984} & Au foil\newline 26.23 mg/cm${}^{2}$ & cyclotron & Faraday cup & Ge(Li) & ${}^{197}$Au(p,3n)${}^{195g}$Hg,$\sigma$, No. 1, 27 MeV\newline ${}^{197}$Au(p,3n)${}^{195m}$Hg,$\sigma$, No. 1, 28 MeV\newline  \\ \hline 
\citep{Wu} & Au foil\newline 200$\sigma$$\mu$m & VdG & Faraday cup & n-Long Counter & ${}^{197}$Au(p,n)${}^{197}$Hg,$\sigma$-rel, No. 1.572-2.594 MeV\newline  \\ \hline 
\citep{Nagame} & Au foil \newline 5-10 mg/cm${}^{2}$ & Cyclotron &  & Ge(Li) & ${}^{197}$Au(p,3n)${}^{195g}$Hg,$\sigma$, No. 17, 20-40.3 MeV\newline ${}^{197}$Au(p,3n)${}^{195m}$Hg,$\sigma$, No. 24, 18-50.8 MeV\newline ${}^{197}$Au(p,3n)${}^{195m}$Hg/${}^{195g}$Hg,$\sigma$${}_{rat}$, No. 14, 20-40.2 MeV\newline ${}^{197}$Au(p,pn)${}^{196g}$Au,$\sigma$, No. 15, 13,9-49 MeV\newline ${}^{197}$Au(p,pn)${}^{196m}$Au,$\sigma$, No. 13, 20.5-49.1 MeV\newline ${}^{197}$Au(p,pn)${}^{196m}$Au/${}^{196g}$Au,$\sigma$${}_{rat}$, No. 13, 20.1-48.9 MeV \\ \hline 

\end{tabular}

\end{center}
\end{table*}

\setcounter{table}{0}
\begin{table*}[t]
\tiny
\caption{continued}
\begin{center}
\begin{tabular}{|p{0.8in}|p{0.7in}|p{0.7in}|p{1.0in}|p{0.8in}|p{1.9in}|} \hline 
\textbf{Author} & \textbf{Target} & \textbf{Irradiation} & \textbf{Beam current\newline measurement\newline and monitor\newline reaction} & \textbf{Measurement \newline of activity} & \textbf{Reaction\newline Measured quantity\newline Number of measured data points \newline Energy range (MeV) } \\ \hline 
\citep{Scholten} & Au foil & Cyclotron\newline Stacked foil & ${}^{nat}$Cu(p,x)${}^{62,65}$Zn\newline ${}^{27}$Al(p,x)${}^{22,24}$Na & Ge(Li) & ${}^{197}$Au(p,x)${}^{7}$Be,$\sigma$,No. 7, 40-92.9 MeV\newline  \\ \hline 
\citep{Szelecsenyi1997} & Au foil\newline 49 mg/cm${}^{2}$\newline 9.4 mg/cm${}^{2}$ & Cyclotron\newline Stacked foil & Faraday & Ge(Li) & ${}^{197}$Au(p,n)${}^{197m}$Hg,$\sigma$, No. 70, 4.7-18.3 MeV\newline ${}^{197}$Au(p,pn)${}^{196}$Au,$\sigma$, No. 24, 13.5-18.3 MeV\newline  \\ \hline 
\citep{Michel} & Au foil\newline  & Cyclotron\newline Syncro-cyclotron\newline Stacked foil & ${}^{nat}$Cu(p,x)${}^{65}$Zn\newline ${}^{27}$Al(p,x)${}^{22}$Na & Ge(Li),HPGe & $^{197}$Au(p,n)$^{197g}$Hg,$\sigma$, No. 7, 43.6-142 MeV\newline ${}^{197}$Au(p,n)${}^{197m}$Hg,$\sigma$, No.8, 28.9-800 MeV\newline ${}^{197}$Au(p,n)${}^{197m}$Hg,$\sigma$, No. 32, 22.3-1600 MeV\newline ${}^{197}$Au(p,pn)${}^{196}$Au,$\sigma$, No. 38, 22.3-2600 MeV\newline ${}^{197}$Au(p,p2n)${}^{195}$Au,$\sigma$, No. 37, 22.3-2600 MeV\newline $^{197}$Au(p,p2n)${}^{194}$Au,$\sigma$, No. 31, 43.6-2600 MeV\newline ${}^{197}$Au(p,pn)${}^{191}$Pt,$\sigma$, No. 6, 40-2600 MeV\newline ${}^{197}$Au(p,pn)$^{192}$Ir,$\sigma$, No. 32, 4-2600 MeV\newline and others \\ \hline 
\citep{Sudar} & Au foil\newline  & Cyclotron\newline Stacked foil & ${}^{63}$Cu(p,n)${}^{63}$Zn\newline ${}^{65}$Zn(p,,n)${}^{65}$Zn & HPGe & ${}^{197}$Au(p,n)$^{197m}$Hg/${}^{197g}$Hg,$\sigma$${}_{rat}$, No.6, 8.8-18.3 MeV \\ \hline 
\citep{Szelecsenyi2008}  & Au foil\newline 4.86 $\mu$m & Cyclotron\newline Stacked foil & Faraday & HPGe & ${}^{197}$Au(p,3n)${}^{195m}$Hg,$\sigma$, No. 20, 25.4-85.4 MeV\newline $^{197}$Au(p,p3n)${}^{194}$Au,$\sigma$, No. 21, 25.4-65.4 MeV\newline ${}^{197}$Au(p,5n)${}^{193m}$Hg,$\sigma$, No. 16, 34.6-65.4 MeV\newline ${}^{197}$Au(p,n)${}^{197m}$Hg,$\sigma$, No. 21, 25.4-65.4 MeV\newline ${}^{197}$Au(p,pn)${}^{196}$Au,$\sigma$, No. 21, 25.4-65.4 MeV \\ \hline 
\citep{Elmaghraby} & Au foil\newline 10 $\mu$m & Cyclotron\newline Stacked foil & ${}^{nat}$Cu(p,x) & HPGe & ${}^{197}$Au(p,n)${}^{197m}$Hg,$\sigma$, No. 10, 5.5-13.9 MeV\newline $^{197}$Au(p,n)${}^{197g}$Hg,$\sigma$, No. 11, 4.5-13.9 MeV \\ \hline 
\citep{Satheesh} & Au foil\newline 3.32 mg/cm$^{2}$ & Cyclotron\newline Stacked foil & Faraday cup & HPGe & ${}^{197}$Au(p,n)${}^{197m}$Hg,$\sigma$, No.7. 8.43-20 MeV\newline $^{197}$Au(p,n)${}^{197g}$Hg,$\sigma$, No. 7, 8.43-20 MeV\newline ${}^{197}$Au(p,n)${}^{197m}$Hg/${}^{197g}$Hg,$\sigma$${}_{rat}$, No. 7, 8.43-20 MeV \\ \hline 
T\'{a}rk\'{a}nyi\newline (2015) this work & Au foil\newline  & Cyclotron\newline Stacked foil & ${}^{27}$Al(p,x)${}^{22,24}$Na\newline ${}^{nat}$Ti(p,x)${}^{48}$V & HPGe & $^{197}$Au(p,n)$^{197m}$Hg,$\sigma$, No.46. 7.24-63.48 MeV\newline ${}^{197}$Au(p,n)${}^{197g}$Hg,$\sigma$, No. 11, 7.25-25.2 MeV\newline ${}^{197}$Au(p,3n)${}^{195m}$Hg,$\sigma$, No.40. 18.74-63.48 MeV\newline ${}^{197}$Au(p,3n)${}^{195g}$Hg,$\sigma$, No.40. 18.74-63.48 MeV\newline ${}^{197}$Au(p,5n)${}^{193m}$Hg,$\sigma$, No.24. 35.46-63.28 MeV\newline $^{197}$Au(p,5n)${}^{193g}$Hg,$\sigma$, No.23. 33.95-63.48 MeV\newline ${}^{197}$Au(p,6n)${}^{192g}$Hg,$\sigma$, No.20. 44.74-63.48 MeV\newline ${}^{197}$Au(p,x)${}^{196m}$Au,$\sigma$, No. 46, 11.97-63.48 MeV\newline ${}^{197}$Au(p,x)${}^{196g}$Au,$\sigma$, No. 48, 7.25-63.48 MeV\newline ${}^{197}$Au(p,x)${}^{195}$Au,$\sigma$, No. 35, 20.88-63.48 MeV\newline $^{197}$Au(p,x)${}^{194}$Au,$\sigma$, No. 44, 7.25-63.48 MeV\newline ${}^{197}$Au(p,x)${}^{191}$Au,$\sigma$, No. 7, 58.08-63.48 MeV\newline ${}^{197}$Au(p,x)${}^{191}$Pt,$\sigma$?$\sigma$No. 19, 37.95-63.48 MeV\newline ${}^{197}$Au(p,x)${}^{192}$Ir,$\sigma$, No. 19, 46.91-63.48 MeV \\ \hline 
\end{tabular}

%\begin{flushleft}
%\tiny{\noindent 
%}
%\end{flushleft}

\end{center}
\end{table*}

\section{Experiment and data evaluation}
\label{3}

The experiment was performed using the activation method, stacked foil irradiation technique and off-line high resolution gamma-ray spectrometry. Cross section data were deduced relative to the chosen monitor reactions, re-measured simultaneously over the whole covered energy range.
Two stacks were irradiated at an external beam line of the Cyclone110 cyclotron of the Université Catholique in Louvain la Neuve (LLN) with a 65 MeV proton beam. The first stack was irradiated at 84 nA for 1 h. It contained a sequence of groups of Al (151.1 $\mu$m), Hf (10.54 $\mu$m), Al (56.6 $\mu$m), Al (26.9 $\mu$m) and GaNi alloy layer (17.7 or 15.9 $\mu$m) electrodeposited on 25 $\mu$m Au foils. The energy range covered by the 18 Au targets was 46.9-63.5 MeV. 
The second stack contained 12 groups of W (21.3$\mu$m), Al (250 $\mu$m) and 13.35 $\mu$m GaNi alloy electrodeposited on 13.35 $\mu$m Cu, Al (250 $\mu$m) followed by 17 blocks of 23.1 $\mu$m Au, 125$\mu$m Al, 21.3 $\mu$m W, 250 $\mu$m Al, 13.2 mm GaNi on 12.5 $\mu$m Cu and 125 $\mu$m Al foils. The energy range of the 17 Au foils was 47.15-23.1 MeV. The stack was irradiated at. 35 nA intensity for 1 h. 
An additional stack was irradiated at the external beam line of VUB CGR 560 cyclotron (Brussels) at 90 nA for 1 hour with a 33.7 MeV incident proton beam. This consisted of 7 groups of Al (11 $\mu$m), In (116 $\mu$m), Al (50 $\mu$m), Ti (10.9 $\mu$m) followed by 11 groups of Au (23.1 $\mu$m), Al (11 $\mu$m), In (116 $\mu$m), Al (50 $\mu$m), Ti (10.9 $\mu$m) The energy range of 11 Au foils was on 7.2 -25.2 MeV.
The targets were mounted in Faraday cup like target holders provided with a long collimator. No chemical separation was used. Gamma-spectra were measured with Canberra HPGe detectors, coupled with plug-in MCA computer card controlled by the Genie2000 software. Four or five series of gamma spectra were measured to follow the decay. The spectra were evaluated  by an iterative method using  the Genie 2000 or Forgamma \citep{Canberra, Szekely} codes.
The used decay data and the Q values of the contributing reactions are collected in Table 2. The decay data were taken from the on-line version NUDAT \citep{NuDat}, the reaction Q-values are obtained from the Q value calculator \citep{Pritychenko}.  
The simultaneously measured excitation functions of the monitor reactions are shown in Fig. 1 in comparison with the recommended data \citep{Tarkanyi2001}. In the high energy irradiations the $^{27}$Al(p,x)$^{22,24}$Na monitor reactions were used, at low energy the $^{nat}$Ti(p,x)$^{48}$V reaction. 
The beam energies in targets (preliminary) were determined by using a home-made code based on the tables and coefficients given in \citep{Andersen} and corrected according to the results of the fitted monitor reactions (final) \citep{Tarkanyi1991}. Uncertainty of energy was obtained taking into account cumulative effects of possible uncertainties (primary energy, target thickness, energy straggling, correction to monitor reaction). 
As Au is monoisotopic, so called isotopic cross sections were determined. Due to the complex gamma spectra, the large number of products and their different half-lives, use of experimental data in literature and of theoretical results (tendency, shape, magnitude) help significantly for data evaluation. Uncertainty of cross sections was determined by taking square root from the sum in quadrature of all individual contributions\citep{Inter}: beam current (7 \%), beam-loss corrections (max. 1.5 \%),  target thickness (1 \%), detector efficiency (5 \%), photo peak area determination  and counting statistics (1-20 %).

\begin{figure}
\includegraphics[width=0.5\textwidth]{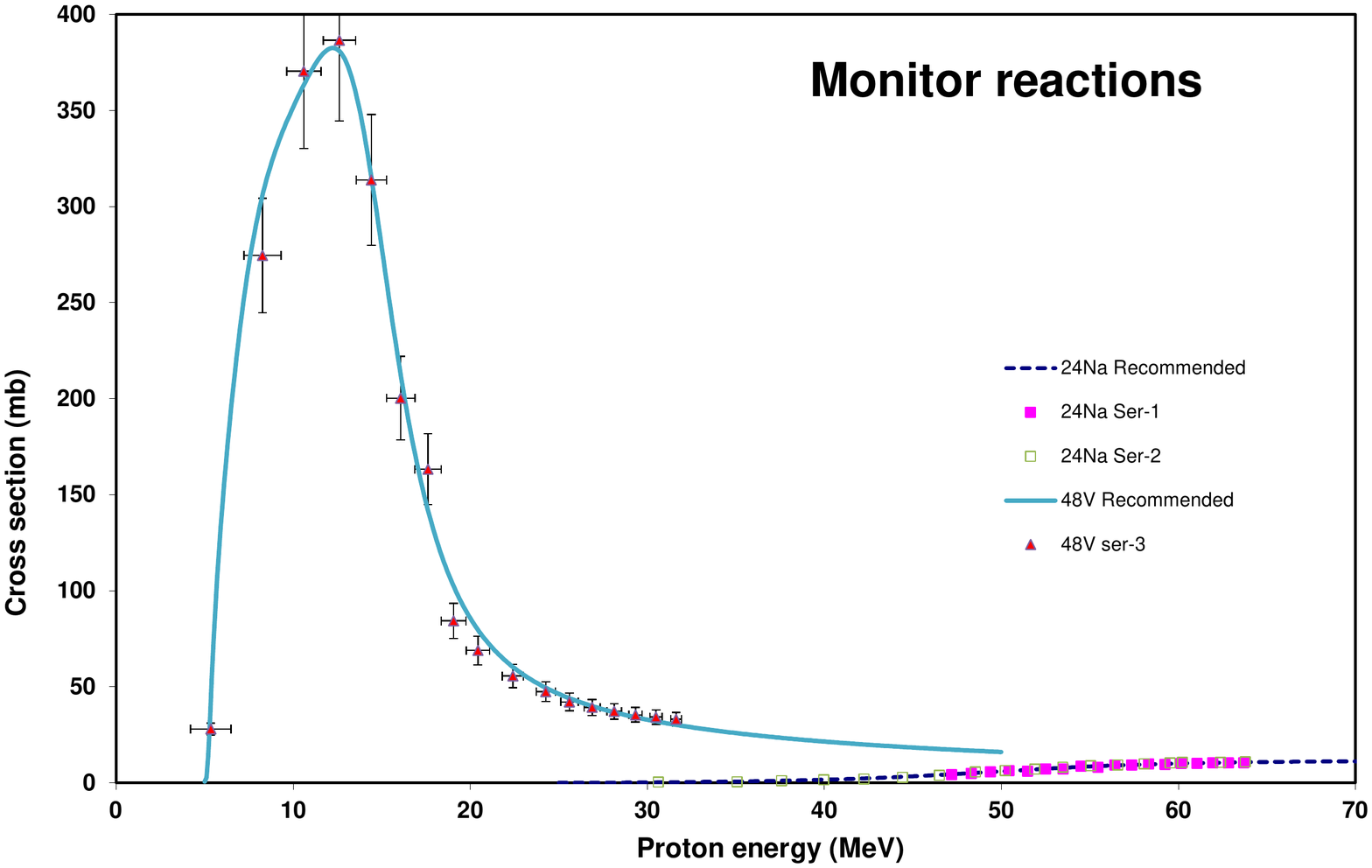}
\caption{Illustration of agreement of simultaneously measured experimental data of $^{27}$Al(p,x)$^{24}$Na and $^{nat}$Ti(p,x)$^{48}$V monitor reactions in comparison with the recommended values to monitor of the proton  beam parameters through the irradiated stacks}
\label{fig:1}       
\end{figure}

\begin{table*}[t]
\tiny
\caption{Decay and nuclear characteristics of the investigated reaction products, contributing reactions and their Q-values}
\begin{center}
\begin{tabular}{|p{0.6in}|p{0.4in}|p{0.7in}|p{0.5in}|p{0.3in}|p{0.9in}|p{0.5in}|} \hline 
\textbf{Nuclide\newline Spin\newline Isomeric level (keV)} & \textbf{Half-life} & \textbf{Decay path (\%)} & \textbf{E$_{\gamma}$(keV)} & \textbf{I$_{\gamma}$(\%)} & \textbf{Contributing process} & \textbf{Q-value\newline (keV)} \\ \hline 
\textbf{${}^{197m}$Hg\newline }13/2${}^{+}$\newline 298.93 & 23.8 h  & EC 8.6\newline IT 91.4 & 133.98\newline 279.01\newline  & 33.5\newline 6 & ${}^{197}$Au(p,n)\newline  & -1681.22\newline  \\ \hline 
\textbf{${}^{197g}$Hg\newline }1/2$^{-}$ & 64.14 h & EC 100\newline  & 77.351\newline 191.437 & 18.7\newline 0.632 & $^{197}$Au(p,n)\newline ${}^{197m}$Hg decay & -1382.29 \\ \hline 
\textbf{${}^{195m}$Hg\newline }13/2$^{+}$\newline 176.07  & 41.6 h & EC 45.8\newline IT 54.2 & 261.75\newline 387.87\newline 560.27 & 31\newline 2.18\newline 7.1 & $^{197}$Au(p,3n)\newline  & -17242.4\newline  \\ \hline 
\textbf{${}^{195g}$Hg\newline }1/2${}^{-}$ & 10.53 h & EC 100\newline  & 180.11\newline 207.1\newline 261.75\newline 585.13\newline 599.66\newline 779.80\newline 1111.04\newline 1172.38 & 1.95\newline 1.6\newline 1.6\newline 2.04\newline 1.83\newline 7.0\newline 1.48\newline 1.28 & $^{197}$Au(p,3n)\newline $^{195m}$Hg decay\newline  & -17066.3 \\ \hline 
\textbf{${}^{193m}$Hg\newline }13/2$^{(+)}$\newline 140.765 & 11.8 h & IT 7.2\newline EC 92 & 257.99\newline 407.63\newline 573.26\newline 932.57 & 49\newline 32\newline 26\newline 12.5 & ${}^{197}$Au(p,5n)\newline  & -33286.5\newline  \\ \hline 
\textbf{${}^{193g}$Hg\newline }3/2${}^{-}$ & 3.8 h & EC 100  & 186.56\newline 381.60\newline 861.11\newline 1118.84 & 15.2\newline 16\newline 12.4\newline 8.0 & $^{197}$Au(p,5n)\newline ${}^{195m}$Hg decay\newline \newline  & -33145.7 \\ \hline 
\textbf{$^{192}$Hg\newline }~0+ & 4.85 h & EC 100  & 157.2\newline 274.8\newline 306.5 & 7.2\newline ~52\newline 5.5 & ${}^{197}$Au(p,6n)\newline  & -40268.8 \\ \hline 
\textbf{$^{196m}$Au\newline }12$^{-}$\newline 595.664 & 9.6 h & IT 100 & 137.69\newline 147.81\newline 168.37\newline 188.27\newline 285.49\newline 316.19 & 1.3\newline 43.5\newline 7.8\newline 30.0\newline 4.4\newline 3.0 & $^{197}$Au(p,pn)\newline  & -8668.02\newline  \\ \hline 
\textbf{$^{196g}$Au\newline }2$^{-}$ & 6.1669 d & EC 93.0\newline $\betaup^{-}$ 7.0 & 333.03\newline 355.73\newline 426.10 & 22.9\newline 87\newline 6.6 & ${}^{197}$Au(p,pn)\newline ${}^{196m}$Au decay\newline  & -8072.36 \\ \hline 
\textbf{$^{195g}$Au\newline }3/2${}^{+}$ & 186.09 d & EC 100 & 98.88\newline 129.757 & 11.2\newline 0.842 & ${}^{197}$Au(p,p2n)\newline ${}^{195m}$Au decay\newline $^{195}$Hg decay & -14714.13 \\ \hline 
\textbf{$^{194}$Au\newline }1$^{-}$ & 38.02 h & EC 100 & 293.548\newline 328.464\newline 1468.882 & 10.58\newline 60.4\newline 6.61 & ${}^{197}$Au(p,p3n)\newline $^{194}$Hg decay\newline  & -23141.77 \\ \hline 
\textbf{$^{191}$Au\newline }1${}^{-}$ & 3.18 h & EC 99.746\newline ${\beta}^{+}$ 0.254 & 277.86\newline 399.84\newline ~586.44\newline 674.22 & 6.4\newline 4.2\newline 15.0\newline 6.0 & $^{197}$Au(p,p6n)\newline $^{191}$Hg decay\newline  & -45757.7 \\ \hline 
\textbf{$^{191}$Pt\newline }3/2$^{-}$\newline  & 2.83 d & $\varepsilon$ 100\newline  & 359.90\newline 409.44\newline 538.90 & 6.0\newline 8.0\newline 13.7 & ${}^{197}$Au(p,2p5n)\newline $^{191}$Au decay & -43085.08 \\ \hline 
\textbf{$^{192}$Ir\newline }~ 4$^{+}$~ & 73.829 d & $\varepsilonup$ 4.76\newline $\betaup^{-}$ 95.24 & 295.95650\newline 308.45507\newline 316.50618\newline 468.06885\newline 604.41105 & 28.71\newline 29.70\newline 82.86\newline 47.84\newline 8.216 & $^{197}$Au(p,3p3n)\newline  & -35095.22 \\ \hline 
\end{tabular}

\begin{flushleft}
\tiny{\noindent When complex particles are emitted instead of individual protons and neutrons, the Q-values have to be decreased by the respective binding energies of the compound particles: np-d: +2.2 MeV; 2np-t: +8.48 MeV; 2p2n-$\alpha$: +28.30 MeV
}
\end{flushleft}

\end{center}
\end{table*}

%\setcounter{table}{0}
%\begin{table*}[t]
%\tiny
%\caption{continued}
%\begin{center}

%\end{center}
%\end{table*}

\section{Comparison with nuclear model calculations}
\label{4}
The cross-sections of the investigated reactions were compared with the data given in the on-line TENDL-2015 \citep{Koning2015} library. This library is based on both default and adjusted TALYS (1.6) calculations \citep{Koning2012}. The calculations we made by EMPIRE 3.2 \citep{Herman2007}(version Malta \citep{Herman2012}) are also presented and compared with the experimental results. In the case of TENDL and EMPIRE the same strategy was followed, i.e. all possible contributions were calculated and added in the final results (if it was possible).

\section{Cross sections}
\label{5}
The experimental cross section data and the comparison with the theoretical results of the TENDL-2014 and TENDL-2015 calculations are shown in Figs. 2-15. The results of the previous version of TENDL is presented in order to demonstrate the improvement in the TALYS calculations. The numerical data are collected in Table 3. The cross section values of mercury radionuclides are due to direct production via (p,xn) reactions. The gold radio-products are produced directly via (p,pxn) reactions and/or additionally through the decay of isobaric parent mercury radioisotopes (cumulative). The ground state of these radioisotopes can additionally be populated through the isomeric transition of the meta-stable state. The cross section is marked with (m+) when the half-life of the isomeric state is significantly shorter than that of the ground state and the cross sections for the production of ground state were deduced from spectra measured after nearly complete decay of the isomeric state.  

\subsection{Production of mercury radioisotopes}
\label{5.1}

\subsubsection{Cross sections for the $^{197}$Au(p,n)$^{197m}$Hg reaction}
\label{5.1.1}
The 23.8 h isomeric state decays for 91.4 \% to the 64.1 h half-life ground state with emission of a 133.98 keV gamma-line and for the rest to stable $^{197}$Au by EC. The production cross sections of the $^{197m}$Hg are shown in Fig. 2, together with the earlier experimental results and theoretical estimations in TENDL-2014 and 2015. There is rather good agreement with earlier experimental data except for the studies of  \citep{Satheesh} and \citep{Hansen}. Unrealistic values for TENDL above 12 MeV can be observed. There are no significant differences between the consecutive TENDL versions. EMPIRE gives much better approximation, especially under 10 MeV and above 20 MeV, it gives also reliable maximum position but overestimates the maximum value.

\begin{figure}
\includegraphics[width=0.5\textwidth]{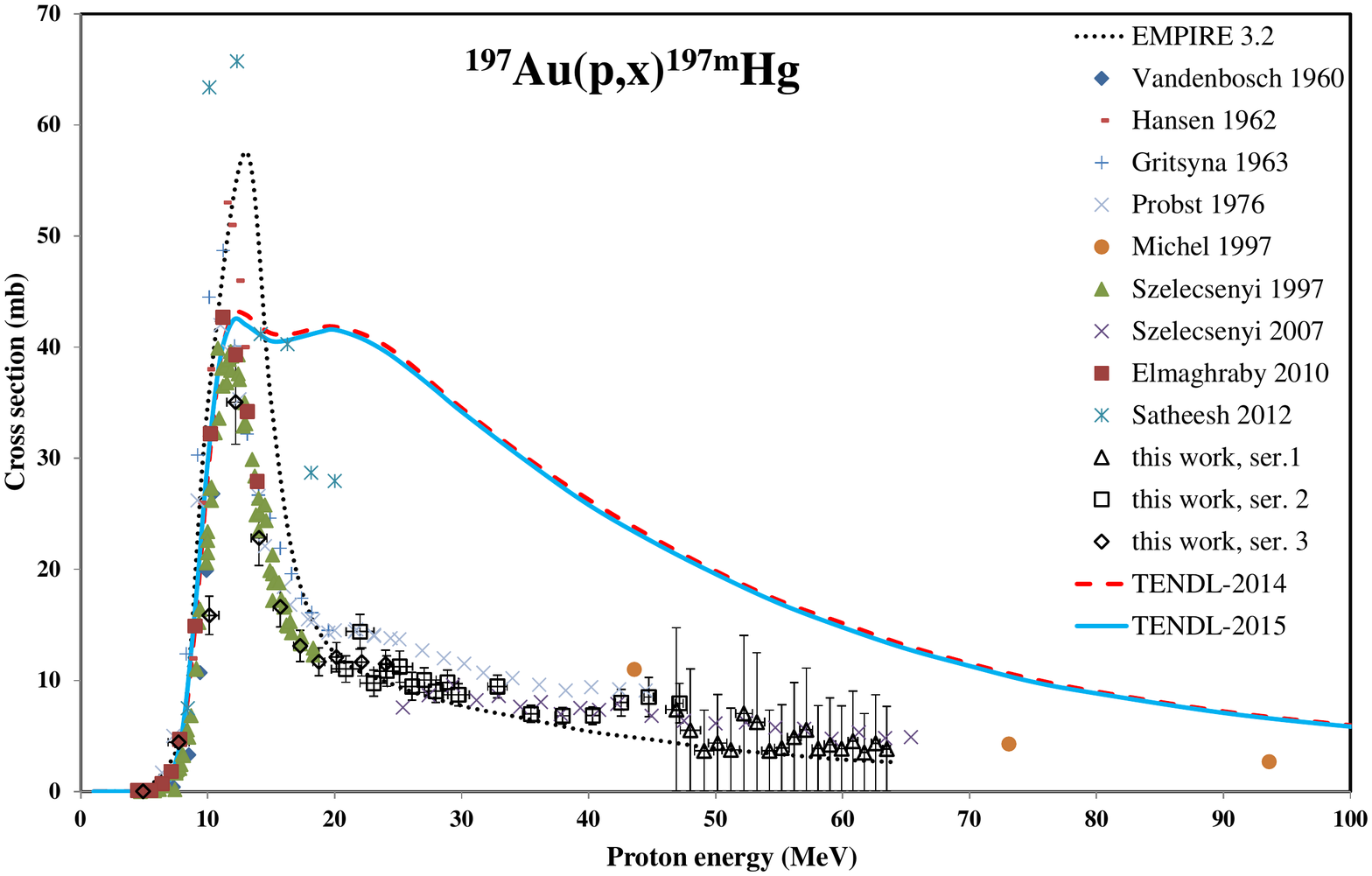}
\caption{Excitation functions of the $^{197}$Au(p,n)$^{197m}$Hg reaction in comparison with literature values and theoretical results from TENDL-2014 and 2015}
\label{fig:2}       
\end{figure}

\subsubsection{Cross sections for the $^{197}$Au(p,n)$^{197g}$Hg reaction}
\label{5.1.2}
The 64.1 h ground state is produced directly via (p,n) reaction and through 91.4 \%  decay of the isomeric state (T$_{1/2}$ = 23.8 h). The experimental data for direct production are shown in Fig. 3. There are large disagreements between the experimental data from the different authors due to several effects: low gamma-energy, weak gamma-lines and separation of the contribution from the decay of isomeric state. The TENDL-2014 and 2015 values seem to give a good description of the shape and the maximum cross section value is near to an average of the experimental ones. The approximation of EMPIRE is better also in this case.

\begin{figure}
\includegraphics[width=0.5\textwidth]{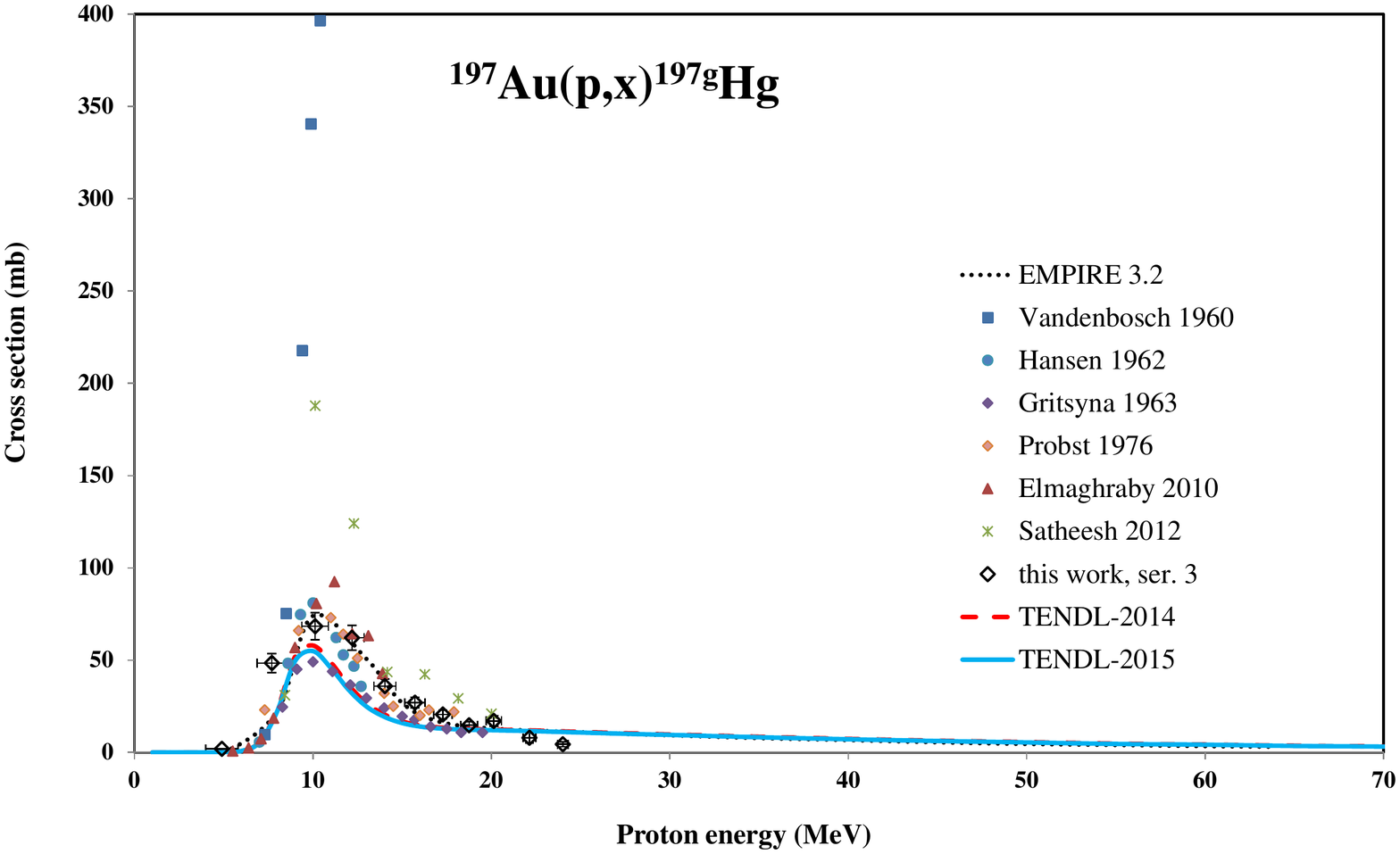}
\caption{Excitation functions of the $^{197}$Au(p,n)$^{197g}$Hg reaction in comparison with literature values and theoretical results from TENDL-2014 and 2015}
\label{fig:3}       
\end{figure}

\subsubsection{Cross sections for the $^{197}$Au(p,3n)$^{195m}$Hg reaction}
\label{5.1.3}
The radionuclide $^{195}$Hg has two states: a longer-lived high spin isomer ($^{195m}$Hg, T$_{1/2}$ = 41.6 h, I$^{\pi}$ = 13/2$^{+}$) and the shorter-lived ground state $^{195g}$Hg (T$_{1/2}$ = 10.53 h, I$^{\pi}$ = 1/2$^{-}$). We obtained production cross sections for both sates. The experimental and the theoretical data for the longer-lived isomeric state are shown in Fig. 4. The agreement is acceptably good between experimental data. The TENDL-2014 and 2015 calculations slightly overestimate the experiments, but the prediction is acceptably good. There is no significant difference between the TENDL versions. Now the overestimation of EMPIRE is strong between 28 and 40 MeV, but its approximation is better at the remaining energy regions.

\begin{figure}
\includegraphics[width=0.5\textwidth]{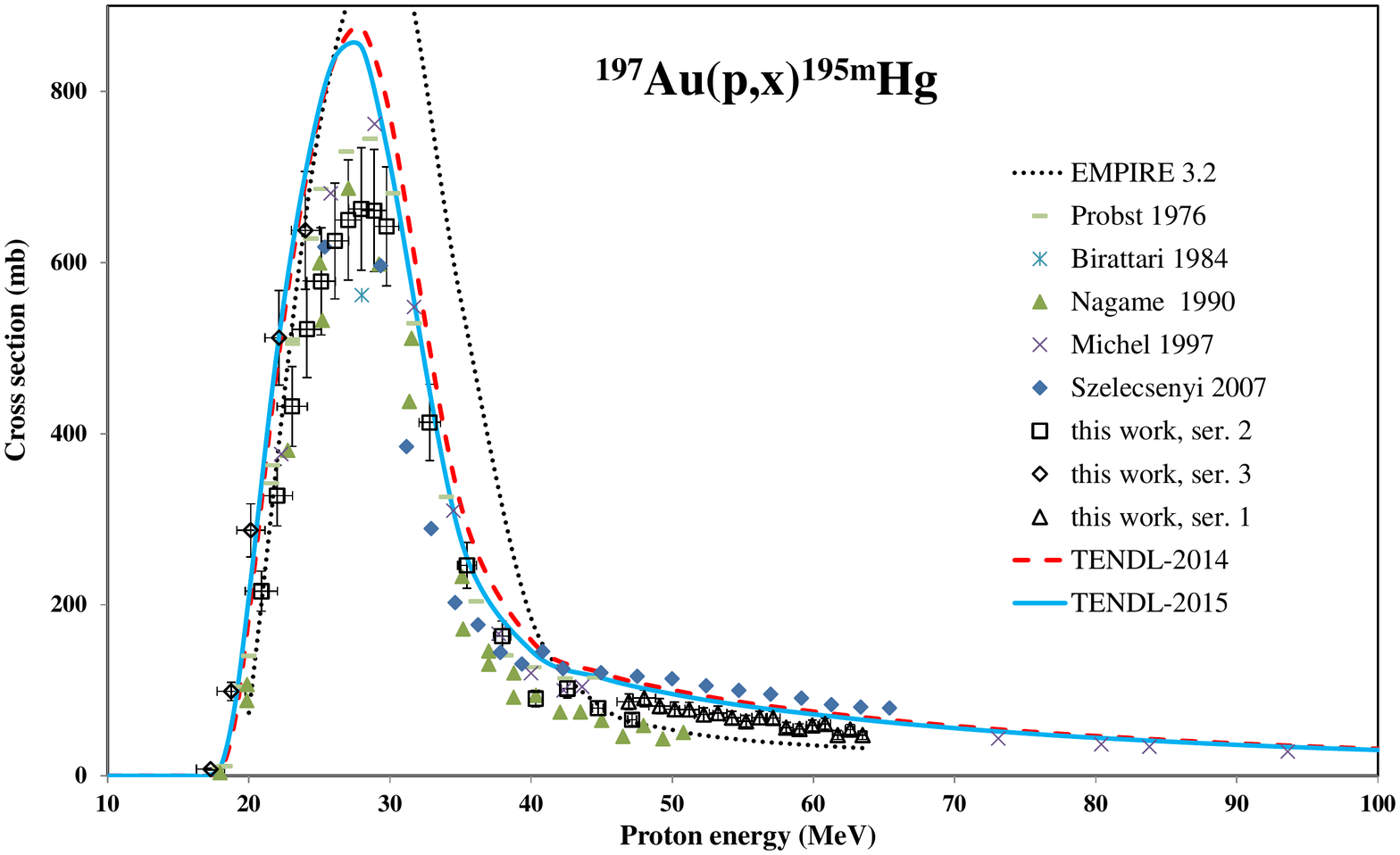}
\caption{Excitation functions of the $^{197}$Au(p,3n)$^{195m}$Hg reaction in comparison with literature values and theoretical results from TENDL-2014 and 2015}
\label{fig:4}       
\end{figure}

\subsubsection{Cross sections for the $^{197}$Au(p,3n)$^{195g}$Hg reaction}
\label{5.1.4}
The 10.53 h ground state is produced directly via (p,n) reaction and through IT 54.2 \% internal  decay of the isomeric state (41.6 h). Figure 5 shows the direct production cross sections obtained after separation of possible contributions of isomeric state decay. The experimental data are very scattered above 40 MeV. The almost identical TENDL versions give good estimation only up to 22 MeV, above this value strongly underestimate and also their maximum energy is shifted downwards. EMPIRE gives better approximation regarding to the maximum position and also the values above it, but underestimates the maximum value similarly to TENDL.

\begin{figure}
\includegraphics[width=0.5\textwidth]{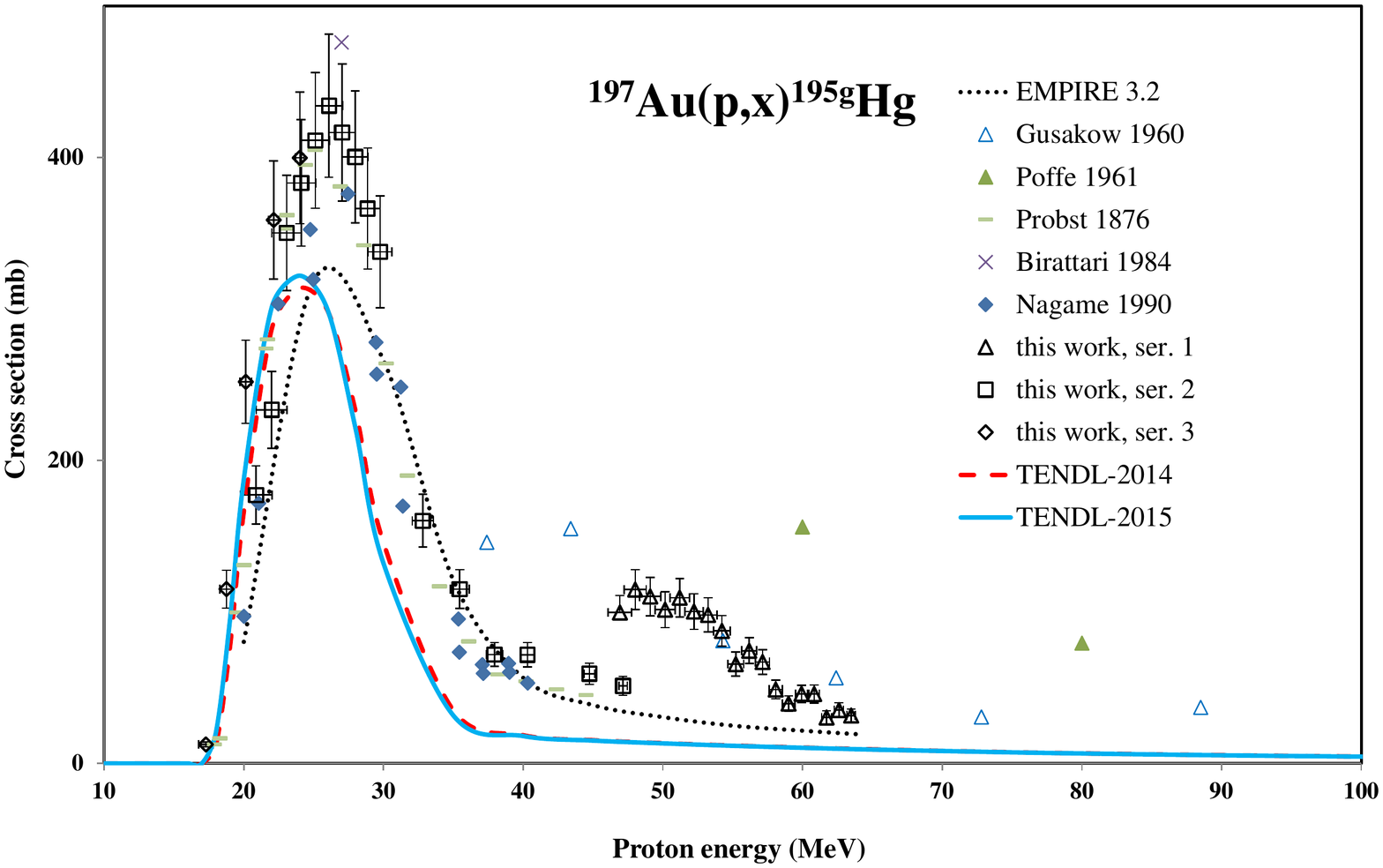}
\caption{Excitation functions of the $^{197}$Au(p,3n)$^{195g}$Hg reaction in comparison with literature values and theoretical results from TENDL-2014 and 2015}
\label{fig:5}       
\end{figure}

\subsubsection{Cross sections for the $^{197}$Au(p,5n)$^{193m}$Hg reaction}
\label{5.1.5}
The radionuclide 193Hg has two states: a longer-lived high spin isomer ($^{193m}$Hg, T$_{1/2}$ = 11.8 h, I$^{\pi}$ = 13/2$^{-}$) and the shorter-lived ground state $^{195g}$Hg (T$_{1/2}$ = 3.8 h, I$^{\pi}$ = 3/2$^{-}$). The experimental and theoretical excitation functions for the metastable state are shown in Fig. 6. Only two earlier experimental data sets were found presenting somewhat lower values confirmed by the TENDL-2014 and 2015 results. In this case EMPIRE is shifted and also overestimates.

\begin{figure}
\includegraphics[width=0.5\textwidth]{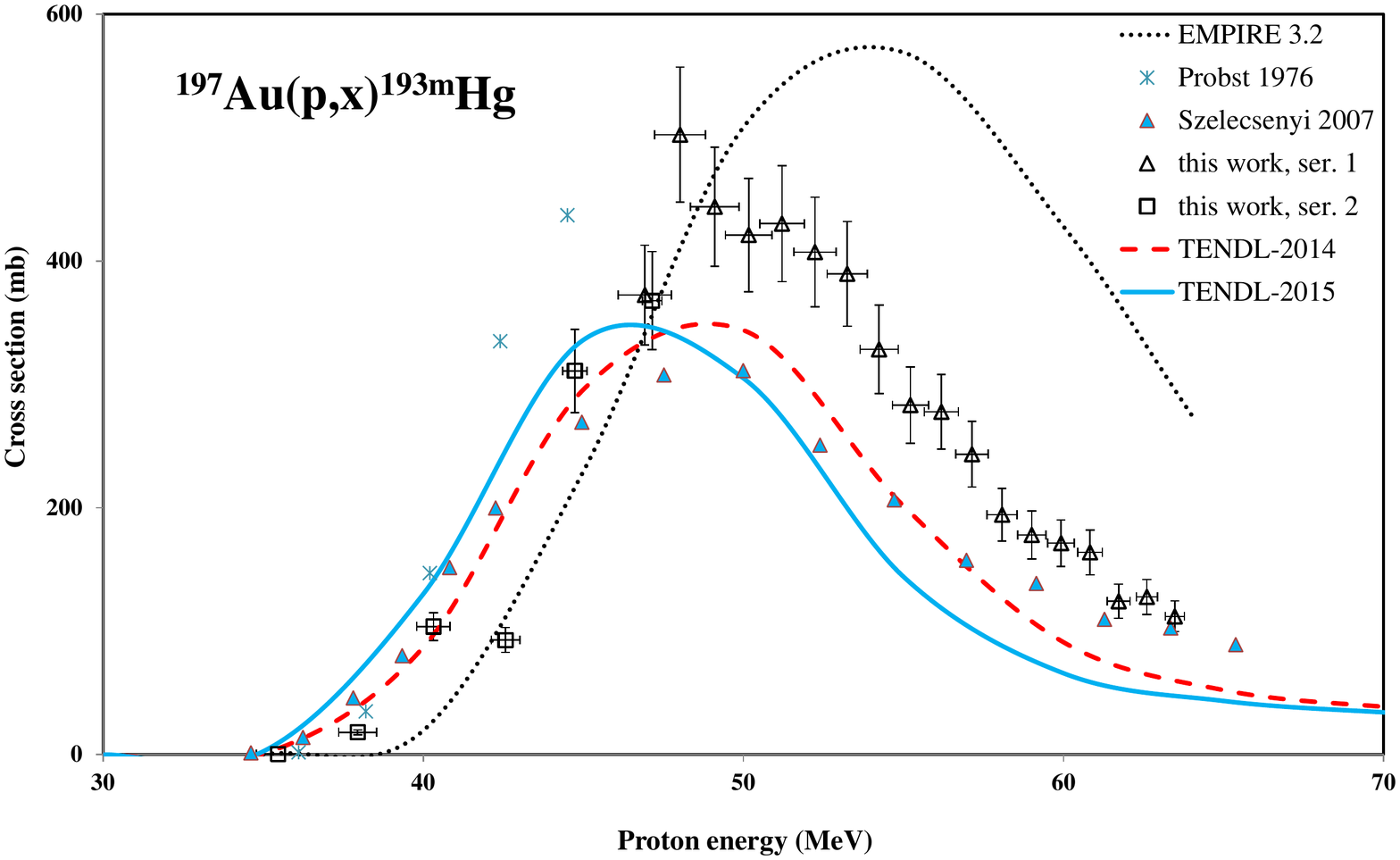}
\caption{Excitation functions of the $^{197}$Au(p,5n)$^{193m}$Hg reaction in comparison with literature values and theoretical results from TENDL-2014 and 2015}
\label{fig:6}       
\end{figure}

\subsubsection{Cross sections for the $^{197}$Au(p,5n)$^{193g}$Hg reaction}
\label{5.1.6}
The experimental excitation function for direct production of $^{193g}$Hg, after correction for contribution from isomeric state decay (IT 7.2 \%) are shown in Fig. 7. Only two earlier experimental data sets were found presenting significantly higher and scattered values. TENDL-2014 and 2015 confirm the shape of our new results but the predicted maximum value is 100 \% higher. In this case the prediction of EMPIRE can be considered as better.

\begin{figure}
\includegraphics[width=0.5\textwidth]{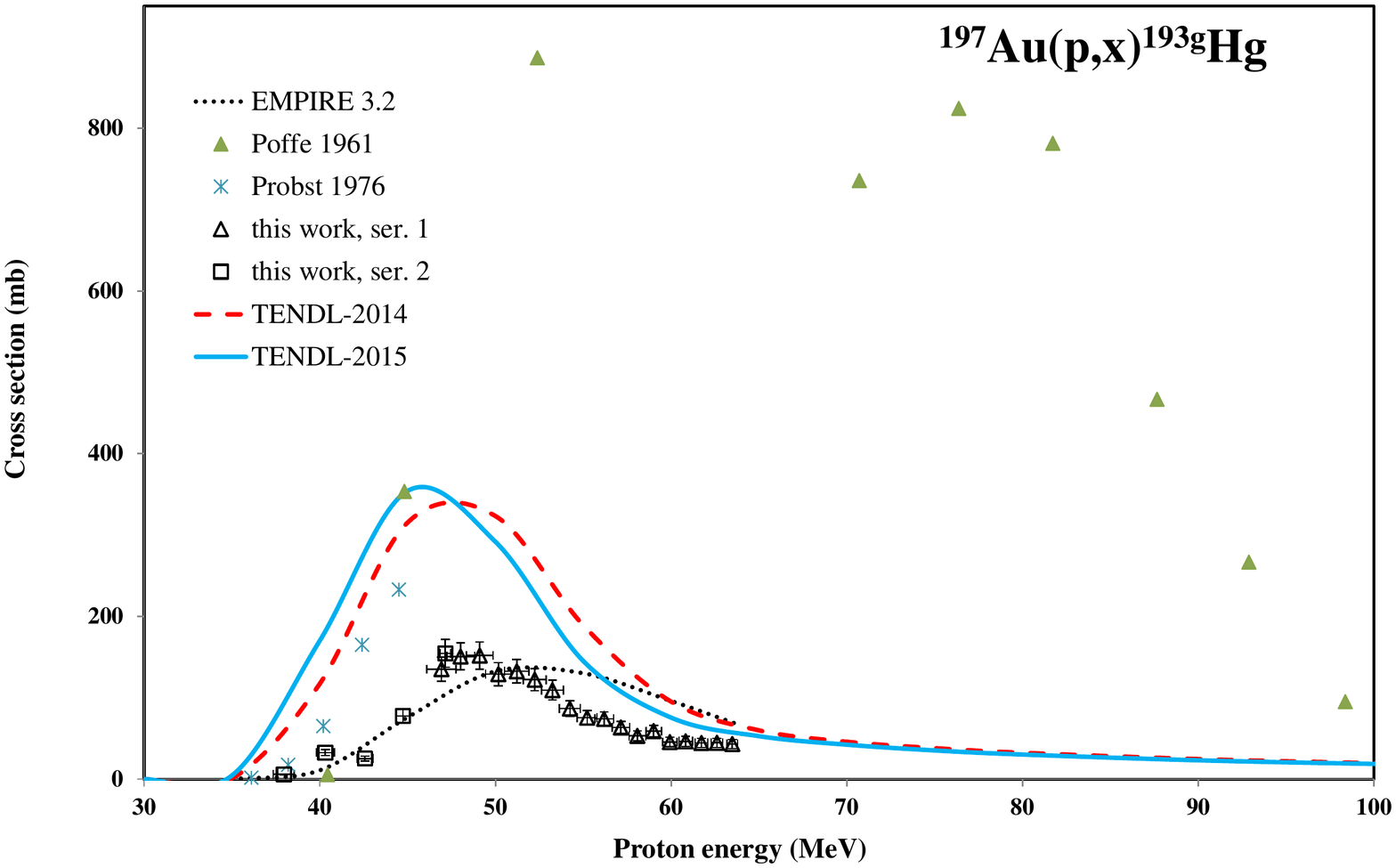}
\caption{Excitation functions of the $^{197}$Au(p,5n)$^{193g}$Hg reaction in comparison with literature values and theoretical results from TENDL-2014 and 2015}
\label{fig:7}       
\end{figure}

\subsubsection{Cross sections for the $^{197}$Au(p,6n)$^{192}$Hg reaction}
\label{5.1.7}
The experimental and the TENDL data for production of $^{192}$Hg (T$_{1/2}$ = 4.85 h) are shown in Fig 8.  Our new data are in good agreement with the TENDL-2014 prediction, but differ significantly from the single literature data. The new TENDL-2015 gives a downwards shift in energy, which is not confirmed by our new experimental results. The EMPIRE code fails to give an acceptable approximation in this case.

\begin{figure}
\includegraphics[width=0.5\textwidth]{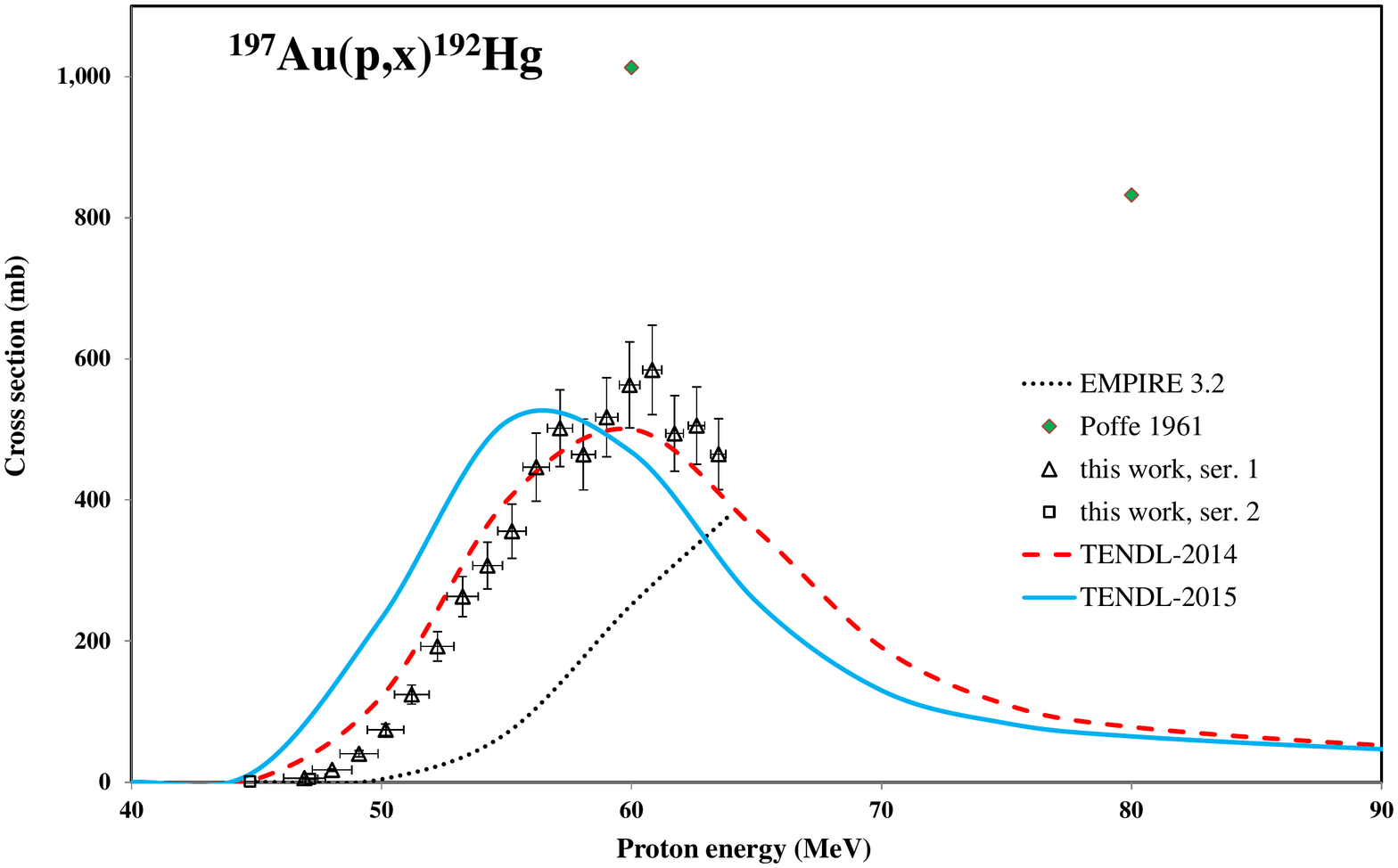}
\caption{Excitation functions of the $^{197}$Au(p,6n)$^{192}$Hg reaction in comparison with literature values and theoretical results from TENDL-2014 and 2015}
\label{fig:8}       
\end{figure}

\subsection{Production of gold radioisotopes}
\label{5.2}
Due to the similar half-lives of the parent Hg radioisotopes and the long cooling time before the first gamma spectra (large contribution from the decay already in the first spectra) we decided not to deduce independent cross sections for production $^{193}$Au and $^{192}$Au (T$_{1/2}$ = 17.65 h and 4.94 h respectively). 

\subsubsection{Cross sections for the $^{197}$Au(p,pn)$^{196m2}$Au reaction}
\label{5.2.1}
We could measure the cross section data for formation of the long-lived, high spin, isomeric state $^{196m2}$Au (T$_{1/2}$ =  9.6 h, 12-, IT 100 \%). No earlier experimental data were found in the literature (Fig. 9). The TENDL-2014 and 2015 values are 6 times higher near the maximum. EMPIRE gives similarly unacceptable result.

\begin{figure}
\includegraphics[width=0.5\textwidth]{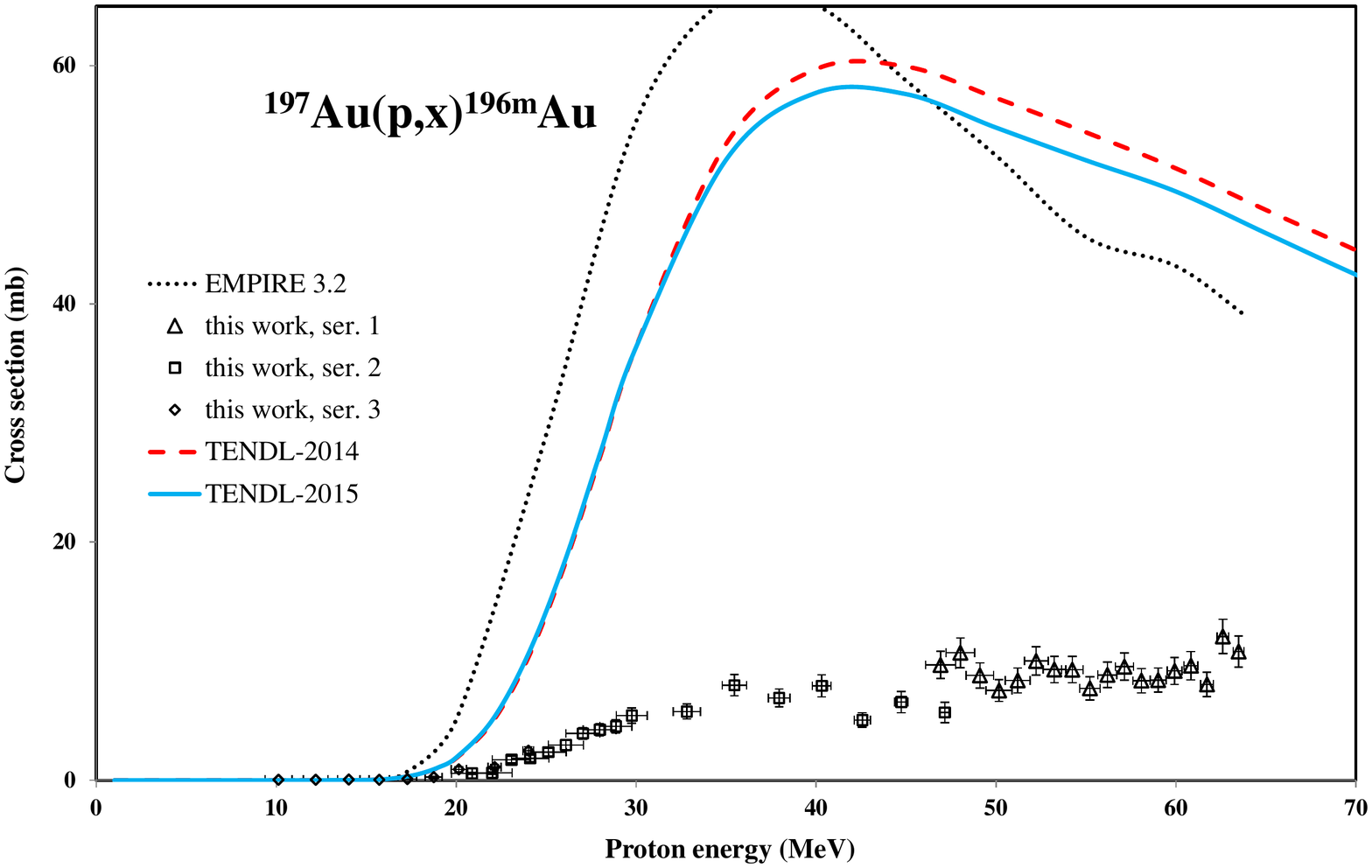}
\caption{Excitation functions of the $^{197}$Au(p,pn)$^{196m2}$Au reaction in comparison with literature values and theoretical results from TENDL-2014 and 2015}
\label{fig:9}       
\end{figure}

\subsubsection{Cross sections for the $^{197}$Au(p,pn)$^{196g}$Au reaction}
\label{5.2.2}
The experimental results for direct production of the $^{196g}$Au ground state (T$_{1/2}$ = 6.1669 d), after correction for contribution of 100 \% isomeric transitions from $^{196m2}$Au but including the total decay of the direct formation of the 8.2 sec $^{197m1}$Au and the 100 \% isomeric transitions from $^{196m2}$Au (T$_{1/2}$ = 9.6 h), are shown in Fig. 10. ($^{196}$Au (cum)). The rather large number of experimental data sets are showing acceptably good agreement. TENDL-2014 and 2015 are slightly higher than all of the experimental values. EMPIRE underestimates in spite of the inclusion of the $^{197m1}$Au contribution.

\begin{figure}
\includegraphics[width=0.5\textwidth]{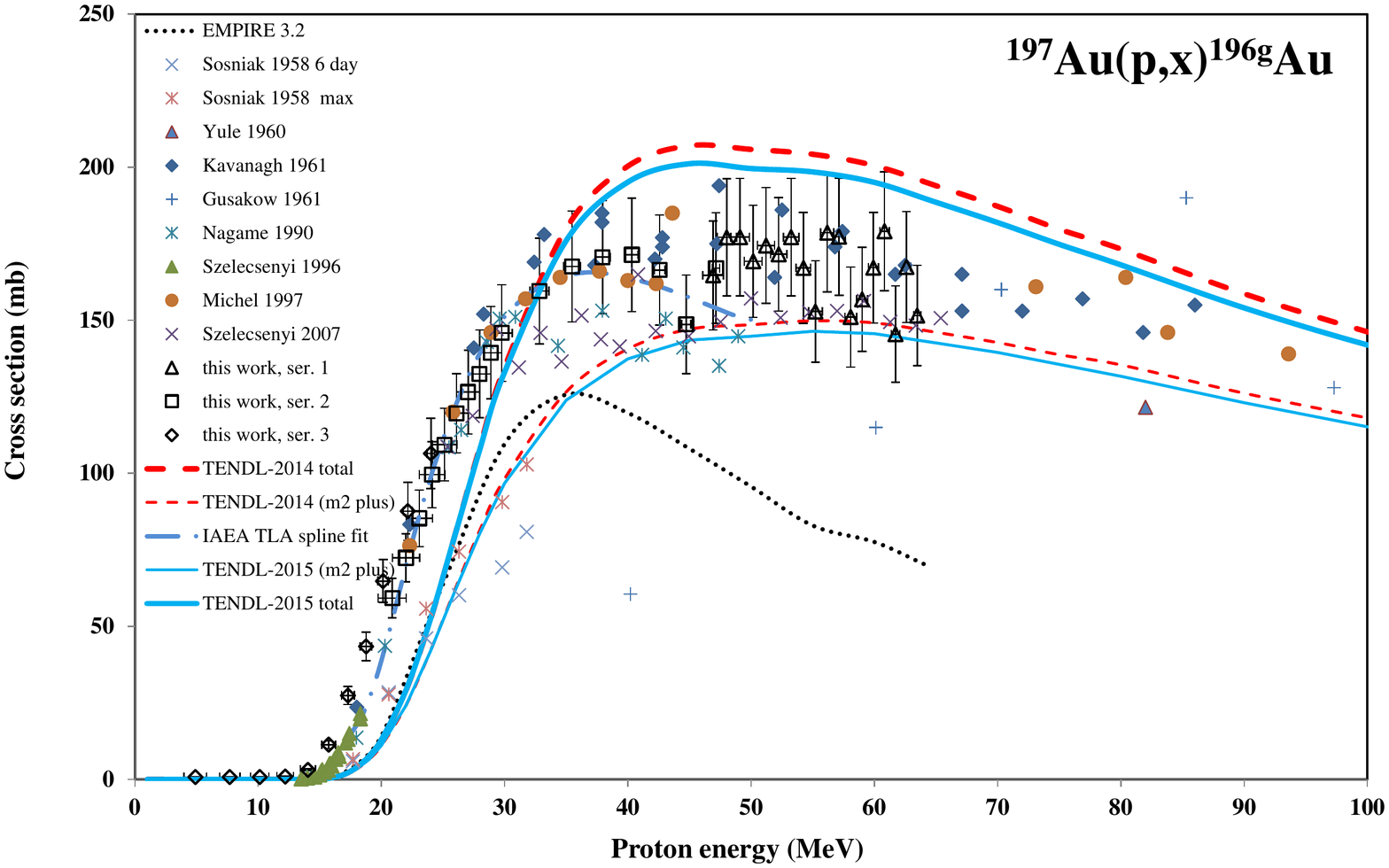}
\caption{Excitation functions of the $^{197}$Au(p,pn)$^{196g}$Au(cum) reaction in comparison with literature values and theoretical results from TENDL-2014 and 2015}
\label{fig:10}       
\end{figure}

\subsubsection{Cross sections for the $^{197}$Au(p,x)$^{195g}$Au reaction}
\label{5.2.3}
We deduced cumulative production cross of $^{195g}$Au (T$_{1/2}$ = 186.09 d), which includes contribution from direct production, from the short-lived $^{195m}$Au isomeric state (T$_{1/2}$ = 30.5 s, IT 100 \%) and from the decay of the isomeric states of parent $^{195m,g}$Hg (T$_{1/2}$ = 41.06h and 10.53 h). Our cross section data for the cumulative production (cum) and the literature results for the direct (g) or cumulative (cum) production of the ground state are shown in Fig. 11. The values of TENDL-2014 and 2015 for cumulative production follow closely the experimental data. EMPIRE prediction is shifted and gives some overestimation between 25 and 45 MeV.

\begin{figure}
\includegraphics[width=0.5\textwidth]{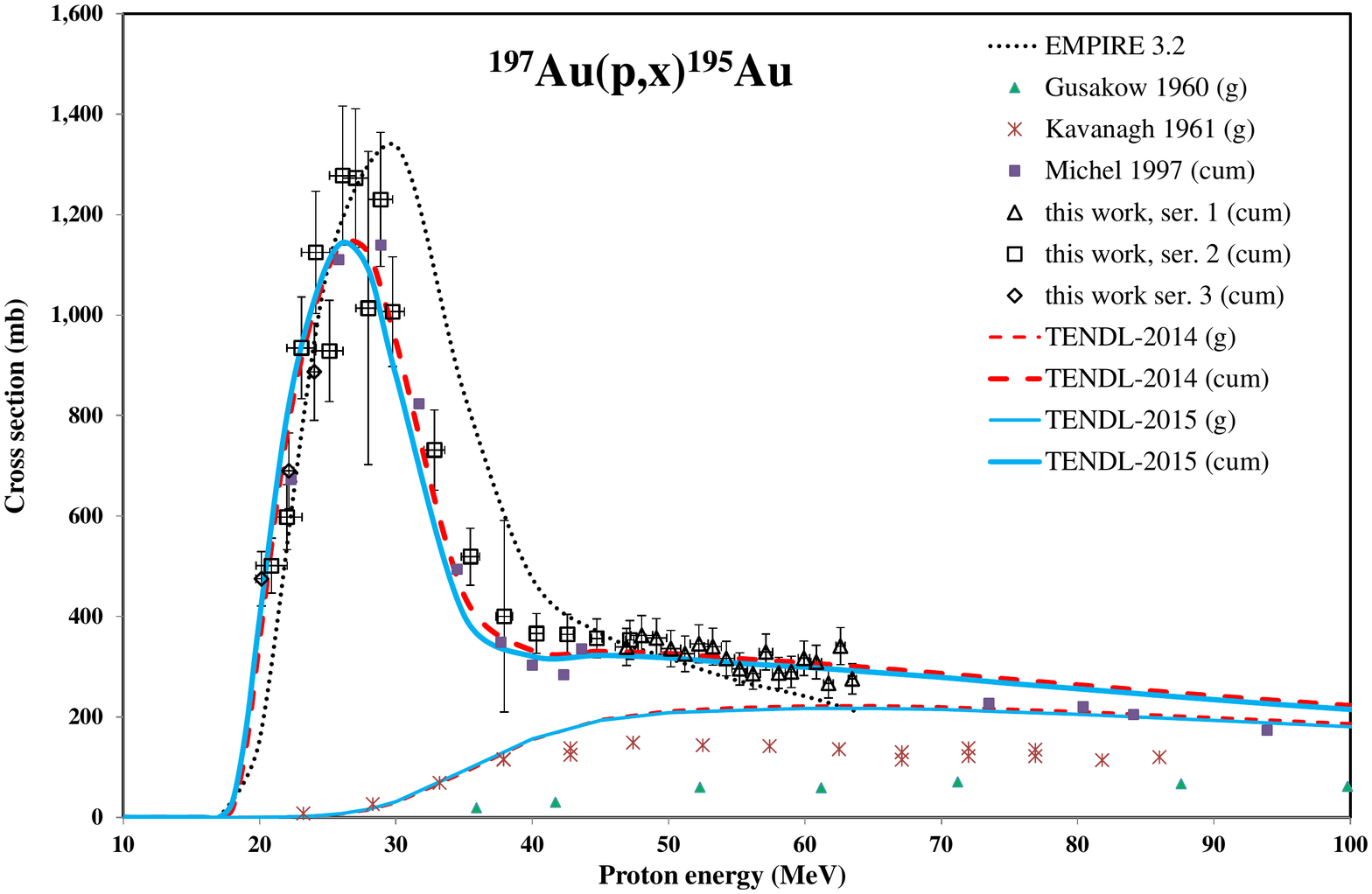}
\caption{Excitation functions of the $^{197}$Au(p,x)$^{195}$Au(cum) and $^{197}$Au(p,x)$^{195g}$Au  reactions in comparison with literature values and theoretical results from TENDL-2014 and 2015}
\label{fig:11}       
\end{figure}

\subsubsection{Cross sections for the $^{197}$Au(p,p3n)$^{194}$Au reaction}
\label{5.2.4}
The direct production cross sections of $^{194}$Au (T$_{1/2}$ = 38.02 h) are shown in Fig. 12 in comparison with the earlier experimental data. On the basis of the theoretical cross section of the parent $^{194}$Hg and taking into account its very long half-life the contribution from the decay of $^{194}$Hg (T$_{1/2}$ = 444 a) in our experimental circumstances and data uncertainties can be neglected. The TENDL-2014 and 2015 predictions follow our experimental results. A strong overestimation can be observed by EMPIRE above 30 MeV.

\begin{figure}
\includegraphics[width=0.5\textwidth]{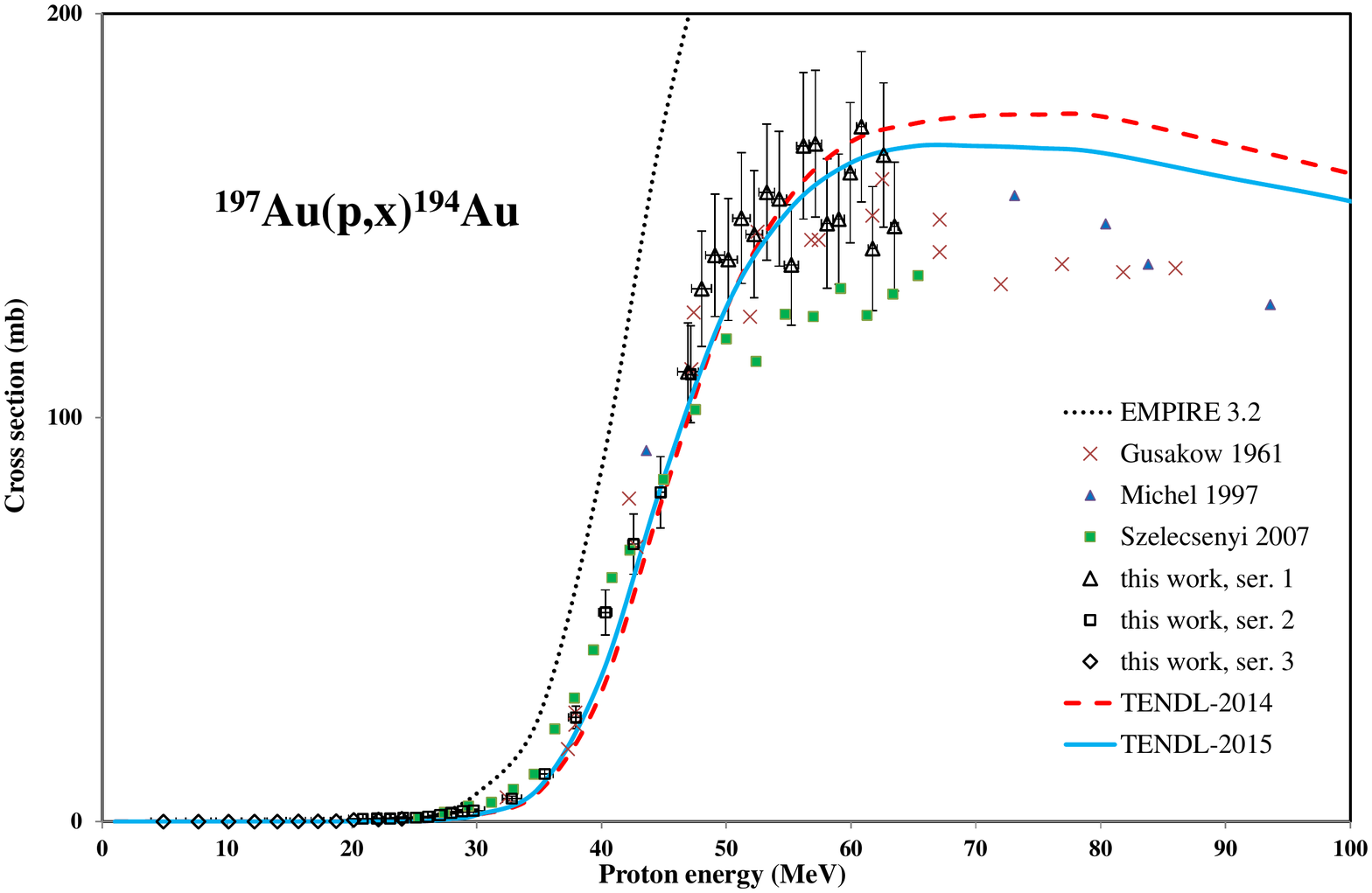}
\caption{Excitation functions of the $^{197}$Au(p,p3n)$^{194}$Au  reaction in comparison with literature values and theoretical results from TENDL-2014 and 2015}
\label{fig:12}       
\end{figure}

\subsubsection{Cross sections for the $^{197}$Au(p,x)$^{191}$Au  reaction}
\label{5.2.5}
The measured cumulative cross sections of $^{191}$Au (T$_{1/2}$ = 3.18 h) contain the direct production and production through the complete decay of the parent $^{191}$Hg (T$_{1/2}$ = 50.58 min) (Fig. 13). No earlier experimental data were found in the literature. Our new data are in good agreement with the TENDL-2014 prediction. In this case the new TENDL-2015 shifted the maximum towards lower energies, which is not confirmed by the experiment. The reason of the strong underestimation of EMPIRE is that the production is mainly fed by the $^{191}$Hg decay, which was not included in the EMPIRE calculation, because of the code failure by large number of emitted particles at higher energies.

\begin{figure}
\includegraphics[width=0.5\textwidth]{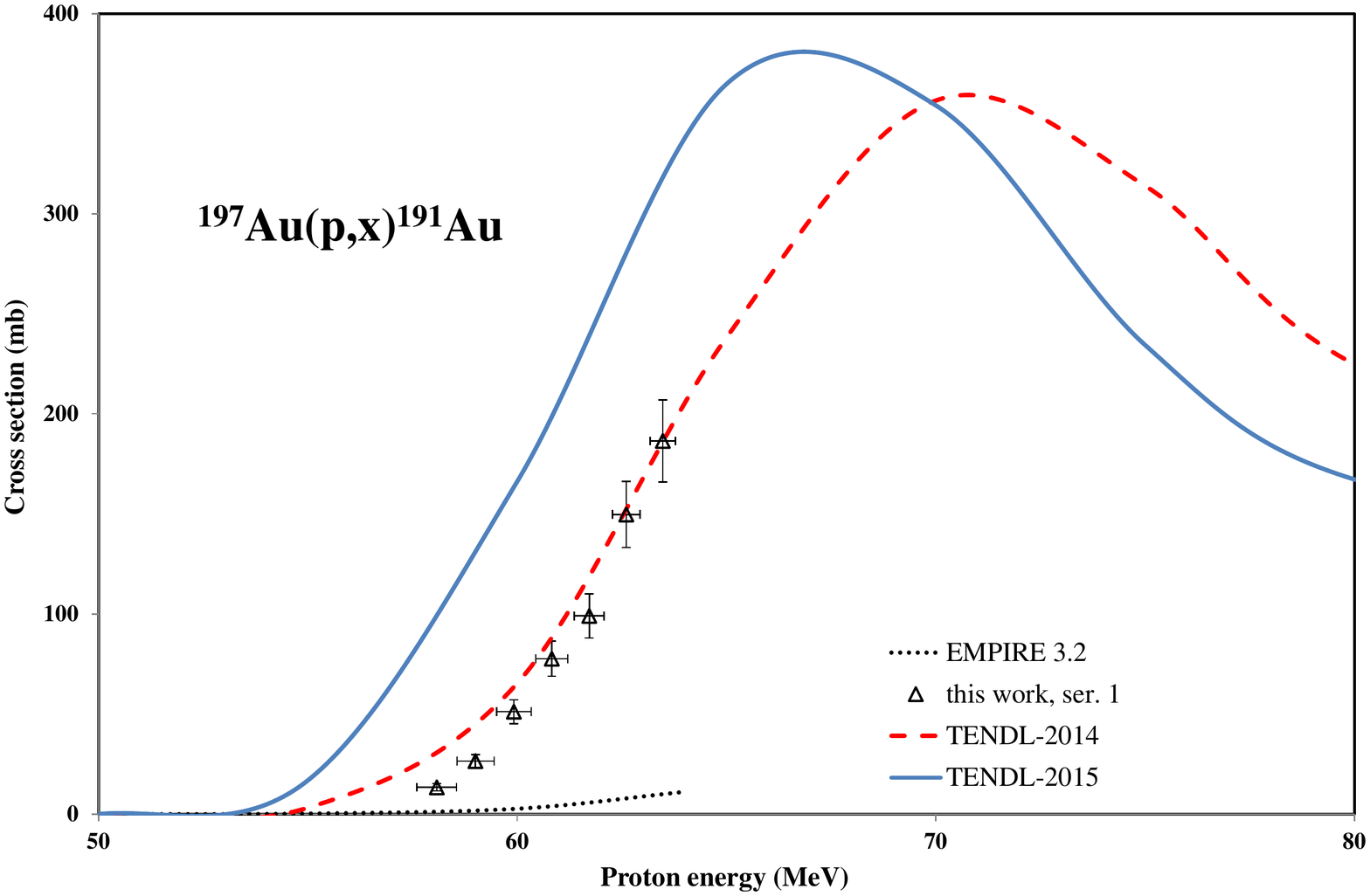}
\caption{Excitation functions of the $^{197}$Au(p,x)$^{191}$Au  reaction in comparison with literature values and theoretical results from TENDL-2014 and 2015}
\label{fig:13}       
\end{figure}

\subsection{Production of platinum radioisotopes}
\label{5.3}

\subsubsection{Cross sections for the $^{197}$Au(p,x)$^{191}$Pt  reaction}
\label{5.3.1}
The measured cumulative cross sections of $^{191}$Pt (T$_{1/2}$ = 2.802 d) (cum = direct $^{191}$Pt + decay of $^{191}$Au + decay of $^{191}$Hg) are practically identical with the production cross sections of $^{191}$Au (Fig. 14, see also Fig. 13), indicating the low contribution from the direct production as confirmed by the TENDL-2014 results. TENDL-2015 is not confirmed by our new experimental results again. No earlier experimental data were found in the literature. The EMPIRE prediction disagrees.

\begin{figure}
\includegraphics[width=0.5\textwidth]{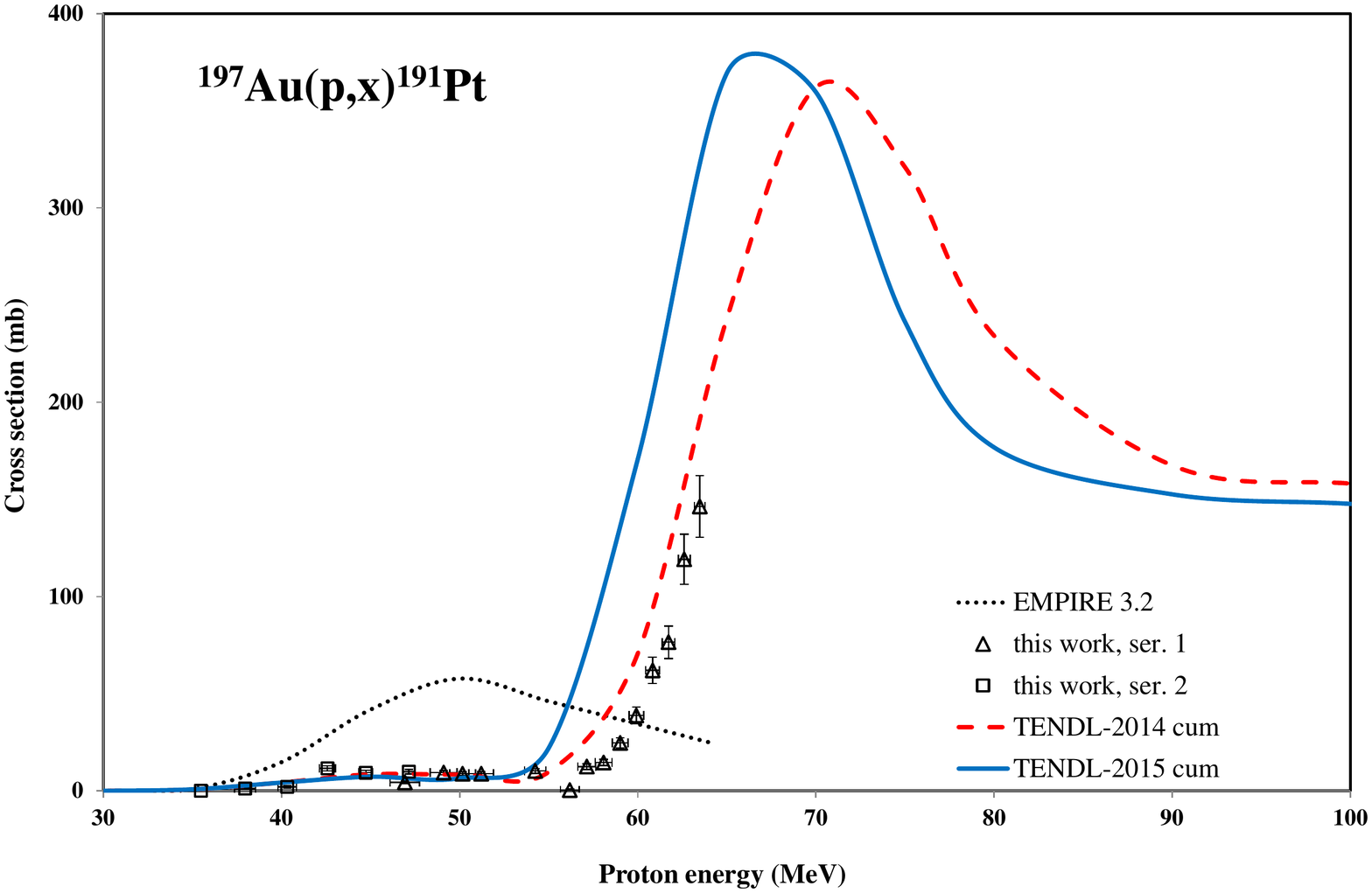}
\caption{Excitation functions of the $^{197}$Au(p,x)$^{191}$Pt(cum)  reaction in comparison with theoretical results from TENDL-2014 and 2015}
\label{fig:14}       
\end{figure}

\subsection{Production of iridium radioisotopes}
\label{5.4}

\subsubsection{Cross sections for the $^{197}$Au(p,x)$^{192}$Ir  reaction}
\label{5.4.1}
The ground state of $^{192}$Ir is a closed radioisotope with three longer-lived isomeric states: a very long-lived, high spin, isomer $^{192m2}$Ir (T$_{1/2}$ = 241 a, I$^{\pi}$ = 11$^-$), a short-lived, low spin, isomeric state $^{192m1}$Ir (T$_{1/2}$ = 1.45 min, I$^{\pi}$ = 1$^-$ ) and the ground state $^{192g}$Ir (T$_{1/2}$ = 73.829 d, I$^{\pi}$ = 4$^+$). The measured cross sections (Fig. 15) contain the direct production and production through the decay of the short half-life isomeric state. Under the used experimental circumstances, the contribution through the decay of the long-lived isomeric does not play a role, so it was neglected. The agreement with the data of  \citep{Michel} is acceptable, the TENDL versions strongly underestimate the experimental values and do not predict the trend of the experimental curves. The EMPIRE strongly overestimates from the threshold.

\begin{figure}
\includegraphics[width=0.5\textwidth]{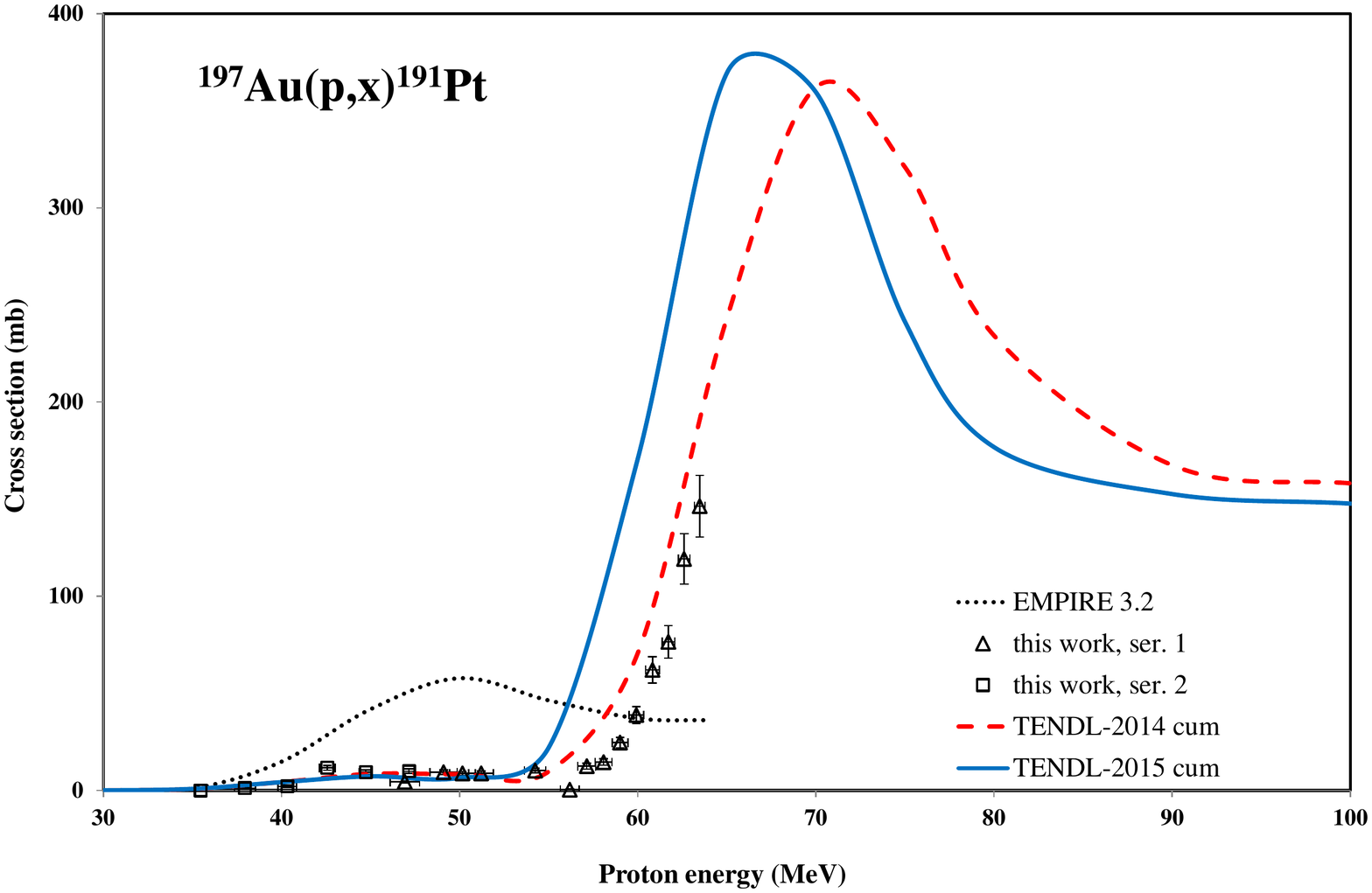}
\caption{Excitation functions of the $^{197}$Au(p,x)$^{192}$Ir(m1+)  reaction in comparison with literature values and theoretical results from TENDL-2014 and 2015}
\label{fig:15}       
\end{figure}

\begin{table*}[t]
\tiny
\caption{Experimental cross sections of $^{197m,195m,195g,193m,193g,192}$Hg, $^{196m,196g(cum),195}$Au(cum), $^{194,191(cum)}$Au, $^{191(cum)}$Pt  and $^{192}$Ir nuclear reactions (series 1, 2 and 3 are separated by thick lines)}
\begin{center}
\begin{tabular}{|p{0.08in}|p{0.08in}|p{0.07in}|p{0.08in}|p{0.07in}|p{0.07in}|p{0.08in}|p{0.08in}|p{0.08in}|p{0.08in}|p{0.08in}|p{0.08in}|p{0.08in}|p{0.08in}|p{0.08in}|p{0.08in}|p{0.08in}|p{0.08in}|p{0.08in}|p{0.08in}|p{0.08in}|p{0.08in}|p{0.08in}|p{0.08in}|p{0.07in}|p{0.07in}|p{0.07in}|p{0.07in}|p{0.07in}|p{0.07in}|} \hline 
\multicolumn{2}{|p{0.1in}|}{\textbf{Energy\newline E$\pm\Delta$E}} & \multicolumn{2}{|c|}{\textbf{${}^{197m}$Hg}} & \multicolumn{2}{|c|}{\textbf{$^{197g}$Hg}} & \multicolumn{2}{|c|}{\textbf{${}^{195m}$Hg}} & \multicolumn{2}{|c|}{\textbf{$^{195g}$Hg}} & \multicolumn{2}{|c|}{\textbf{$^{193m}$Hg}} & \multicolumn{2}{|c|}{\textbf{$^{193g}$Hg}} & \multicolumn{2}{|c|}{\textbf{$^{192}$Hg}} & \multicolumn{2}{|c|}{\textbf{$^{196m}$Au\newline cum}} & \multicolumn{2}{|c|}{\textbf{$^{196g}$Au}} & \multicolumn{2}{|c|}{\textbf{$^{195}$Au\newline cum}} & \multicolumn{2}{|c|}{\textbf{$^{194}$Au}} & \multicolumn{2}{|c|}{\textbf{$^{191}$Au\newline cum}} & \multicolumn{2}{|c|}{\textbf{$^{191m}$Pt}} & \multicolumn{2}{|c|}{\textbf{$^{192}$Ir}} \\ \hline 
\multicolumn{2}{|p{0.1in}|}{\textbf{MeV}} & \multicolumn{28}{|c|}{\textbf{Cross section $\sigma \pm \Delta\sigma$ (mb)}} \\ \hline 
63.5 & 0.3 & 3.8 & 0.5 & ~ &  & 47.6 & 5.2 & 31.6 & 4.2 & 112.0 & 12.4 & 43.1 & 5.7 & 464.8 & 50.3 & 10.8 & 1.3 & 151.6 & 16.4 & 275.7 & 29.9 & 147.3 & 15.9 & 186.5 & 20.5 & 146.4 & 15.9 & 0.41 & 0.05 \\ \hline 
62.6 & 0.3 & 4.4 & 0.8 & ~ &  & 54.2 & 5.9 & 35.2 & 4.7 & 127.7 & 14.2 & 44.6 & 5.7 & 505.4 & 54.7 & 12.1 & 1.4 & 167.4 & 18.1 & 341.0 & 37.0 & 165.0 & 17.9 & 149.7 & 16.5 & 119.2 & 13.0 & 0.37 & 0.04 \\ \hline 
61.7 & 0.4 & 3.5 & 0.6 & ~ &  & 47.6 & 5.2 & 30.4 & 4.4 & 124.3 & 13.9 & 44.5 & 5.6 & 494.4 & 53.5 & 8.1 & 1.0 & 145.5 & 15.7 & 266.9 & 28.9 & 141.9 & 15.4 & 99.0 & 11.0 & 76.5 & 8.3 & 0.35 & 0.04 \\ \hline 
60.8 & 0.4 & 4.5 & 1.0 & ~ &  & 60.7 & 6.6 & 45.9 & 6.0 & 163.8 & 18.2 & 46.5 & 6.2 & 584.4 & 63.2 & 9.6 & 1.2 & 179.1 & 19.4 & 309.4 & 33.5 & 172.0 & 18.6 & 77.7 & 8.8 & 62.1 & 6.8 & 0.39 & 0.04 \\ \hline 
59.9 & 0.4 & 3.9 & 0.8 & ~ &  & 58.8 & 6.4 & 46.0 & 5.6 & 171.3 & 18.8 & 45.7 & 5.8 & 563.1 & 60.9 & 9.2 & 1.1 & 167.2 & 18.1 & 316.9 & 34.3 & 160.6 & 17.4 & 51.1 & 6.0 & 38.8 & 4.3 & 0.38 & 0.04 \\ \hline 
59.0 & 0.4 & 4.2 & 0.5 & ~ &  & 54.7 & 5.9 & 39.5 & 5.0 & 177.9 & 19.5 & 59.0 & 6.8 & 517.4 & 56.0 & 8.4 & 1.0 & 156.8 & 17.0 & 289.7 & 31.4 & 149.1 & 16.1 & 26.5 & 3.3 & 24.6 & 2.7 & 0.38 & 0.04 \\ \hline 
58.1 & 0.5 & 3.9 & 1.0 & ~ &  & 56.3 & 6.1 & 48.8 & 6.2 & 194.4 & 21.4 & 53.2 & 6.1 & 464.5 & 50.3 & 8.4 & 1.0 & 151.1 & 16.4 & 287.7 & 31.2 & 148.0 & 16.0 & 13.4 & 1.8 & 14.6 & 1.6 & 0.32 & 0.04 \\ \hline 
57.1 & 0.5 & 5.6 & 1.1 & ~ &  & 67.8 & 7.3 & 67.0 & 8.1 & 243.4 & 26.7 & 63.5 & 7.4 & 501.7 & 54.3 & 9.5 & 1.1 & 177.3 & 19.2 & 329.4 & 35.7 & 167.8 & 18.2 & ~ &  & 12.5 & 1.5 & 0.33 & 0.04 \\ \hline 
56.2 & 0.5 & 4.9 & 1.0 & ~ &  & 68.1 & 7.4 & 74.4 & 8.6 & 277.8 & 30.3 & 74.2 & 8.3 & 446.7 & 48.3 & 8.9 & 1.1 & 178.6 & 19.3 & 286.3 & 31.0 & 167.3 & 18.1 & ~ &  & 0.4 & 0.5 & 0.25 & 0.03 \\ \hline 
55.2 & 0.6 & 3.9 & 0.5 & ~ &  & 63.8 & 6.9 & 65.6 & 7.9 & 283.1 & 30.9 & 75.6 & 8.6 & 355.9 & 38.5 & 7.7 & 1.0 & 152.9 & 16.5 & 295.8 & 32.1 & 137.8 & 14.9 & ~ &  &  &  & 0.30 & 0.04 \\ \hline 
54.2 & 0.6 & 3.7 & 1.1 & ~ &  & 67.8 & 7.3 & 87.5 & 10.1 & 328.4 & 35.8 & 86.8 & 9.7 & 307.0 & 33.2 & 9.3 & 1.1 & 167.1 & 18.1 & 316.3 & 34.3 & 154.2 & 16.7 & ~ &  & 10.2 & 1.2 & 0.29 & 0.03 \\ \hline 
53.2 & 0.6 & 6.2 & 0.7 & ~ &  & 73.4 & 7.9 & 98.0 & 11.3 & 389.5 & 42.4 & 109.4 & 12.2 & 263.2 & 28.5 & 9.3 & 1.1 & 177.1 & 19.2 & 340.1 & 36.8 & 155.8 & 16.9 & ~ &  & ~ &  & 0.27 & 0.03 \\ \hline 
52.2 & 0.7 & 7.0 & 0.8 & ~ &  & 71.6 & 7.7 & 100.2 & 11.8 & 407.2 & 44.4 & 122.2 & 13.6 & 192.4 & 20.8 & 10.0 & 1.2 & 171.6 & 18.6 & 345.9 & 37.5 & 145.4 & 15.7 & ~ &  & ~ &  & 0.24 & 0.03 \\ \hline 
51.2 & 0.7 & 3.7 & 1.0 & ~ &  & 77.4 & 8.4 & 109.3 & 12.6 & 430.2 & 46.9 & 132.6 & 14.6 & 124.3 & 13.5 & 8.4 & 1.0 & 174.4 & 18.9 & 325.7 & 35.3 & 149.4 & 16.2 & ~ &  & 8.8 & 1.1 & 0.24 & 0.03 \\ \hline 
50.2 & 0.7 & 4.4 & 0.9 & ~ &  & 77.9 & 8.4 & 101.4 & 11.8 & 420.9 & 45.9 & 129.1 & 14.2 & 74.2 & 8.0 & 7.5 & 0.9 & 169.3 & 18.3 & 336.0 & 36.4 & 139.1 & 15.1 & ~ &  & 8.8 & 1.0 & 0.18 & 0.02 \\ \hline 
49.1 & 0.8 & 3.7 & 0.8 & ~ &  & 81.6 & 8.8 & 110.1 & 12.6 & 443.9 & 48.3 & 152.0 & 16.7 & 40.3 & 4.4 & 8.8 & 1.1 & 177.1 & 19.2 & 357.2 & 38.7 & 140.2 & 15.2 & ~ &  & 9.4 & 1.1 & 0.17 & 0.02 \\ \hline 
48.0 & 0.8 & 5.5 & 0.7 & ~ &  & 90.7 & 9.8 & 114.7 & 13.1 & 502.3 & 54.8 & 150.8 & 16.6 & 17.2 & 1.9 & 10.7 & 1.3 & 177.1 & 19.2 & 362.4 & 39.3 & 131.9 & 14.3 & ~ &  & ~ &  & 0.17 & 0.02 \\ \hline 
46.9 & 0.8 & 7.4 & 0.9 & ~ &  & 86.4 & 9.4 & 99.7 & 11.1 & 372.3 & 40.4 & 135.2 & 14.9 & 5.7 & 0.7 & 9.7 & 1.1 & 164.7 & 17.8 & 339.3 & 36.8 & 111.4 & 12.1 & ~ &  & 4.4 & 0.5 & 0.16 & 0.02 \\ \hline  \hline
47.1 & 0.3 & 8.0 & 1.8 & ~ & ~ & 65.4 & 7.1 & 51.2 & 6.2 & 367.8 & 39.8 & 154.7 & 17.2 & 4.7 & 0.5 & 5.7 & 0.9 & 167.0 & 18.1 & 353.3 & 38.5 & 110.7 & 12.0 & ~ & ~ & 9.9 & 1.1 & 0.12 & 0.03 \\ \hline 
44.7 & 0.4 & 8.5 & 1.8 & ~ &  & 79.0 & 8.6 & 59.1 & 7.1 & 310.8 & 33.7 & 77.6 & 8.8 & 1.0 & 0.2 & 6.6 & 0.9 & 148.7 & 16.1 & 356.4 & 38.9 & 81.5 & 8.8 & ~ &  & 9.4 & 1.2 & ~ &  \\ \hline 
42.6 & 0.4 & 8.0 & 1.2 & ~ &  & 101.9 & 11.0 & ~ &  & 92.8 & 10.1 & 25.4 & 3.0 & ~ &  & 5.1 & 0.6 & 166.4 & 18.0 & 364.2 & 40.0 & 68.7 & 7.4 & ~ &  & 11.6 & 1.3 & ~ &  \\ \hline 
40.3 & 0.5 & 6.8 & 0.8 & ~ &  & 90.2 & 9.8 & 71.8 & 8.1 & 103.8 & 11.3 & 32.8 & 3.7 & ~ &  & 7.9 & 0.9 & 171.4 & 18.5 & 366.2 & 39.8 & 51.8 & 5.6 & ~ &  & 2.1 & 0.6 & ~ &  \\ \hline 
37.9 & 0.6 & 6.8 & 0.7 & ~ &  & 163.1 & 17.7 & 71.9 & 7.9 & 18.0 & 2.0 & 5.9 & 0.7 & ~ &  & 6.9 & 0.8 & 170.6 & 18.5 & 400.4 & 190.7 & 25.8 & 2.8 & ~ &  & 1.3 & 0.2 & ~ &  \\ \hline 
35.5 & 0.7 & 7.0 & 0.8 & ~ &  & 246.0 & 26.6 & 115.1 & 12.7 & 0.5 & 0.1 & ~ &  & ~ &  & 8.0 & 0.9 & 167.6 & 18.1 & 519.1 & 56.5 & 11.8 & 1.3 & ~ &  & ~ &  & ~ &  \\ \hline 
32.8 & 0.8 & 9.4 & 1.0 & ~ &  & 413.1 & 44.7 & 160.2 & 17.4 & ~ &  & ~ &  & ~ &  & 5.8 & 0.6 & 159.6 & 17.3 & 731.2 & 79.7 & 5.7 & 0.6 & ~ &  & ~ &  & ~ &  \\ \hline 
29.8 & 0.9 & 8.7 & 1.0 & ~ &  & 642.3 & 69.5 & 337.7 & 36.9 & ~ &  & ~ &  & ~ &  & 5.4 & 0.6 & 145.8 & 15.8 & 1006.8 & 109.3 & 2.7 & 0.3 & ~ &  & ~ &  & ~ &  \\ \hline 
28.9 & 0.9 & 9.8 & 1.1 & ~ &  & 660.9 & 71.5 & 366.3 & 39.9 & ~ &  & ~ &  & ~ &  & 4.5 & 0.5 & 139.4 & 15.1 & 1230.3 & 133.3 & 2.6 & 0.3 & ~ &  & ~ &  & ~ &  \\ \hline 
28.0 & 0.9 & 9.0 & 1.0 & ~ &  & 662.7 & 71.7 & 400.4 & 43.5 & ~ &  & ~ &  & ~ &  & 4.3 & 0.5 & 132.5 & 14.3 & 1013.9 & 311.8 & 2.1 & 0.2 & ~ &  & ~ &  & ~ &  \\ \hline 
27.0 & 0.9 & 10.0 & 1.1 & ~ &  & 649.8 & 70.3 & 416.5 & 45.3 & ~ &  & ~ &  & ~ &  & 3.9 & 0.5 & 126.5 & 13.7 & 1272.7 & 137.9 & 1.6 & 0.2 & ~ &  & ~ &  & ~ &  \\ \hline 
26.1 & 1.0 & 9.5 & 1.3 & ~ &  & 625.2 & 67.6 & 434.3 & 47.3 & ~ &  & ~ &  & ~ &  & 3.0 & 0.4 & 119.6 & 12.9 & 1277.7 & 138.4 & 1.2 & 0.2 & ~ &  & ~ &  & ~ &  \\ \hline 
25.1 & 1.0 & 11.2 & 1.4 & ~ &  & 578.1 & 62.5 & 411.2 & 44.8 & ~ &  & ~ &  & ~ &  & 2.4 & 0.3 & 109.4 & 11.8 & 928.7 & 100.9 & 1.0 & 0.1 & ~ &  & ~ &  & ~ &  \\ \hline 
24.1 & 1.0 & 10.9 & 1.4 & ~ &  & 522.0 & 56.5 & 383.2 & 41.8 & ~ &  & ~ &  & ~ &  & 1.9 & 0.3 & 99.5 & 10.8 & 1124.9 & 121.8 & 1.0 & 0.1 & ~ &  & ~ &  & ~ &  \\ \hline 
23.1 & 1.1 & 9.8 & 1.2 & ~ &  & 431.7 & 46.7 & 350.2 & 38.2 & ~ &  & ~ &  & ~ &  & 1.7 & 0.3 & 85.3 & 9.2 & 934.6 & 101.2 & 0.74 & 0.09 & ~ &  & ~ &  & ~ &  \\ \hline 
22.0 & 1.1 & 14.4 & 1.6 & ~ &  & 327.5 & 35.4 & 233.4 & 25.3 & ~ &  & ~ &  & ~ &  & 0.6 & 0.1 & 72.4 & 7.8 & 597.9 & 64.9 & 0.76 & 0.08 & ~ &  & ~ &  & ~ &  \\ \hline 
20.9 & 1.1 & 11.0 & 1.2 & ~ &  & 215.8 & 23.4 & 177.2 & 19.2 & ~ &  & ~ &  & ~ &  & 0.6 & 0.1 & 59.2 & 6.4 & 501.1 & 54.5 & 0.71 & 0.09 & ~ &  & ~ &  & ~ &  \\ \hline \hline
24.0 & 0.3 & 11.5 & 1.2 & 4.3 & 1.3 & 637.6 & 69.0 & 399.7 & 43.4 & ~ & ~ & ~ & ~ & ~ & ~ & 2.5 & 0.3 & 106.5 & 11.5 & 887.0 & 96.5 & 0.72 & 0.08 & ~ & ~ & ~ & ~ & ~ & ~ \\ \hline 
22.1 & 0.4 & 11.7 & 1.3 & 7.9 & 2.9 & 512.1 & 55.4 & 358.8 & 38.9 & ~ &  & ~ &  & ~ &  & 1.1 & 0.2 & 87.6 & 9.5 & 690.1 & 75.5 & 0.54 & 0.07 & ~ &  & ~ &  & ~ &  \\ \hline 
20.1 & 0.4 & 12.1 & 1.3 & 17.0 & 2.9 & 286.9 & 31.1 & 251.9 & 27.4 & ~ &  & ~ &  & ~ &  & 0.91 & 0.20 & 64.8 & 7.0 & 474.9 & 54.0 & 0.43 & 0.06 & ~ &  & ~ &  & ~ &  \\ \hline 
18.8 & 0.5 & 11.7 & 1.3 & 14.6 & 2.0 & 98.6 & 10.7 & 115.0 & 12.5 & ~ &  & ~ &  & ~ &  & 0.26 & 0.08 & 43.4 & 4.7 & ~ &  & 0.14 & 0.02 & ~ &  & ~ &  & ~ &  \\ \hline 
17.3 & 0.5 & 13.1 & 1.4 & 20.5 & 2.3 & 7.8 & 0.9 & 12.5 & 1.5 & ~ &  & ~ &  & ~ &  & 0.10 & 0.04 & 27.4 & 3.0 & ~ &  & 0.07 & 0.01 & ~ &  & ~ &  & ~ &  \\ \hline 
15.7 & 0.6 & 16.6 & 1.8 & 27.0 & 2.9 & ~ &  & ~ &  & ~ &  & ~ &  & ~ &  & 0.05 & 0.01 & 11.4 & 1.2 & ~ &  & 0.06 & 0.01 & ~ &  & ~ &  & ~ &  \\ \hline 
14.0 & 0.6 & 22.8 & 2.5 & 35.9 & 3.9 & ~ &  & ~ &  & ~ &  & ~ &  & ~ &  & 0.06 & 0.01 & 3.2 & 0.3 & ~ &  & 0.04 & 0.005 & ~ &  & ~ &  & ~ &  \\ \hline 
12.2 & 0.7 & 35.1 & 3.8 & 62.1 & 6.7 & ~ &  & ~ &  & ~ &  & ~ &  & ~ &  & 0.05 & 0.01 & 0.93 & 0.10 & ~ &  & 0.03 & 0.004 & ~ &  & ~ &  & ~ &  \\ \hline 
10.1 & 0.7 & 15.9 & 1.7 & 68.3 & 7.3 & ~ &  & ~ &  & ~ &  & ~ &  & ~ &  & 0.04 & 0.01 & 0.81 & 0.09 & ~ &  & 0.01 & 0.002 & ~ &  & ~ &  & ~ &  \\ \hline 
7.7 & 0.8 & 4.4 & 0.5 & 48.4 & 5.2 & ~ &  & ~ &  & ~ &  & ~ &  & ~ &  & ~ &  & 0.80 & 0.09 & ~ &  & ~ &  & ~ &  & ~ &  & ~ &  \\ \hline 
4.9 & 0.9 & 0.02 & 0.002 & 1.9 & 0.2 & ~ &  & ~ &  & ~ &  & ~ &  & ~ &  & ~ &  & 0.76 & 0.08 & ~ &  & ~ &  & ~ &  & ~ &  & ~ &  \\ \hline 
\end{tabular}

\end{center}
\end{table*}

\section{Integral yields}
\label{6}

From excitation functions integral thick target yields, obtained by Spline fit to our experimental cross section data, were calculated and shown in Fig. 16 as a function of the energy for the radionuclides with known applications ($^{197m,g}$Hg, $^{195m,g}$Hg and $^{196m,g}$Au). Experimental thick target yields exist in the literature \citep{Abe, Birattari, Dmitriev1983, Dmitriev1981}, which are also presented in Fig. 16. For $^{197m}$Hg our results are slightly lower than the results of \citep{Abe} and much larger than the single data point of \citep{Birattari} at 16 MeV. In the case of $^{197g}$Hg the single data point of \citep{Dmitriev1981} is slightly larger than our value, while the curve given by Birattari et al. is larger and the single data point of Abe et al. is lower than our values. In the case $^{195m}$Hg the data of Birattari et al. run slightly under our results, above 20 MeV the agreement is rather acceptable. The results for the ground state $^{195g}$Hg show similar behavior as the metastable state. In the case of $^{196m}$Au the previous data of Birattari et al. are above our results, while by the ground state the agreements with the single points of Dmitriev and Molin at 22 MeV and Abe et al. at 16 MeV as well as with the curve from Birattari et al. are excellent. 

\begin{figure}
\includegraphics[width=0.5\textwidth]{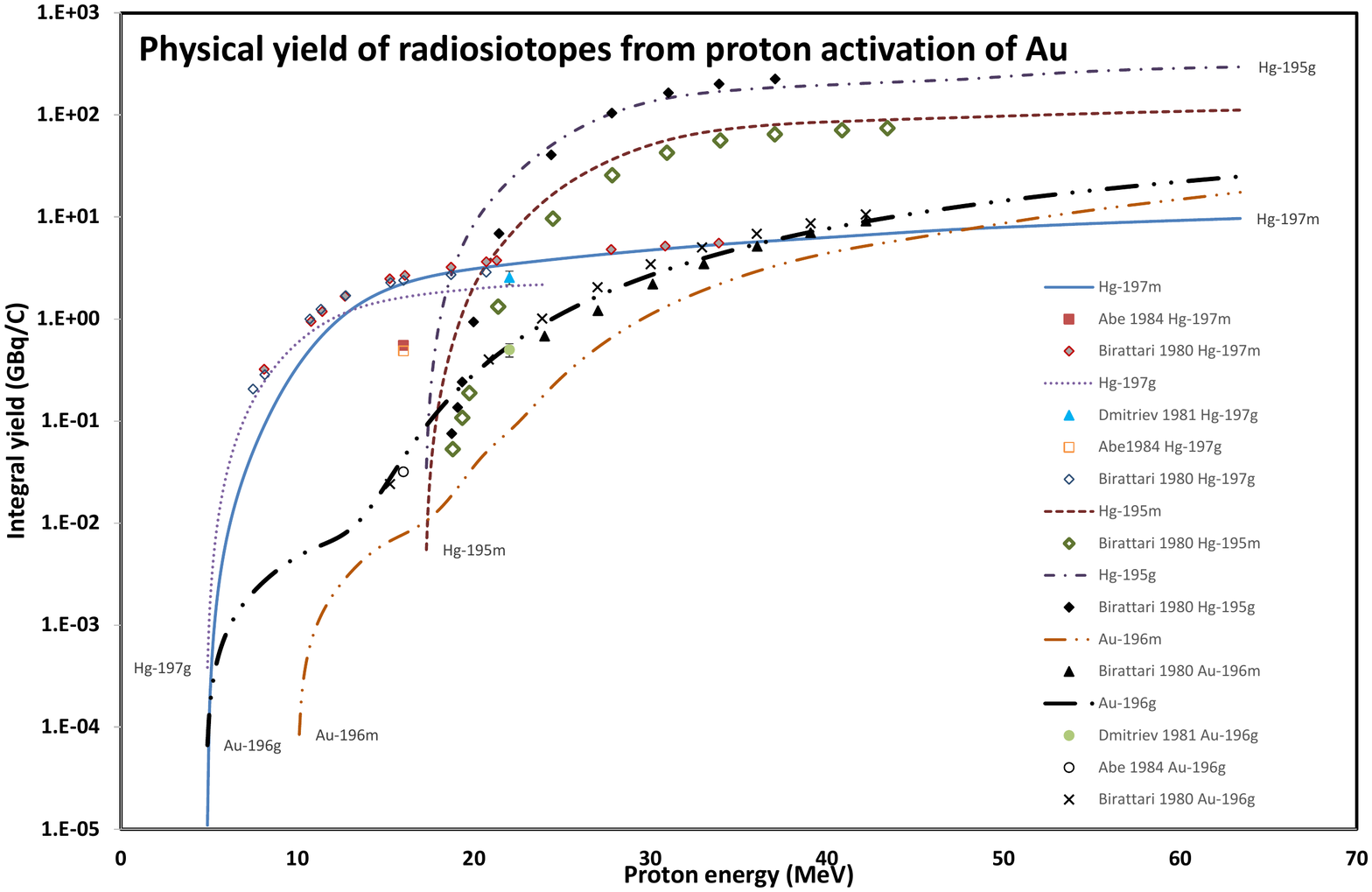}
\caption{Integral yields calculated from the measured excitation functions compared with the literature data}
\label{fig:16}       
\end{figure}

\section{Applications}
\label{7}
Possible medical applications for therapeutic nuclear medicine were recently discussed in detail in our works on activation cross sections of deuteron induced reactions on gold published recently \citep{Tarkanyi2011, Tarkanyi2015}. Here we discuss only the relevance for applications in the thin layer activation technique (TLA) and for beam monitoring.

\subsection{Thin layer activation}
\label{7.1}
For TLA investigation of specimens containing metallic gold or gold alloys, only the longer-lived $^{198g}$Au (T$_{1/2}$ = 2.7 d) and $^{196g}$Au (T$_{1/2}$ = 6.2 d) are suitable among the reaction products studied here and in our previous work \citep{Tarkanyi2015}. Application of $^{196g}$Au was already presented in detail in the TLA data library of the IAEA \citep{IAEA-NDS}. We compare the IAEA recommended data with our new results in Fig. 10. The comparison shows that the experimental data should be refitted, especially above 45 MeV.

\subsection{Beam monitoring}
\label{7.2}
A thin layer of gold is frequently used as target backing material due to its favorable physical and chemical properties. When irradiating the target a few reactions are induced simultaneously in the gold-backing resulting in radioproducts suitable for beam energy and intensity determination. The quality of the gold cross section data however is still not satisfactory. Among the investigated reactions the reactions resulting in $^{197m}$Hg (above 10 MeV), $^{195m}$Hg (above 20 MeV), $^{196g}$Au (above 20 MeV), $^{195}$Au (above 20 MeV, long measurement is necessary) and $^{194}$Au (above 40 MeV) could be used (see Figures 2, 4, 10, 11 and 12).

\section{Summary and conclusion}
\label{8}

Excitation functions of $^{197m,197g,195m,195g,193m,193g,192}$Hg, $^{196m,196g(cum),195g(cum),194,191(cum)}$Au, $^{191(cum)}$Pt  and $^{192}$Ir nuclear reactions are reported  up to 65 MeV, some of them for the first time, relative to well documented monitor reactions. Detailed compilation of earlier experimental data was performed. The agreement is acceptable except for a few reactions. The TENDL-2014 describes well the experiments except for a few isomeric states, while TENDL-2015 shows a little worse prediction than the previous TENDL version. The EMPIRE codes gives better predictions mainly in the cases of a few particle emissions, but fails when complex emissions are also possible and at higher energies. The extended experimental data base provides a basis for improved model calculations and for applications in radioisotope production, in accelerator technology, in charged particle activation analysis and in thin layer application.

\section{Acknowledgements}
\label{}
This work was done in the frame of MTA-FWO (Vlaanderen) research projects. The authors acknowledge the support of research projects and of their respective institutions in providing the materials and the facilities for this work. One of the authors (F. Ditrói) would also acknowledge the support of IAEA and namely to R. Capote and M. Herman for the valuable advices in installing and using the newest EMPIRE code for charged particle reactions.
%\FloatBarrier
 


%% References

%\clearpage
\bibliographystyle{elsarticle-harv}
\bibliography{Aup}






% \end{thebibliography}

\end{document}